\documentclass[10pt]{article}

\pdfoutput=1

\usepackage{amsmath,amssymb,amsfonts,amscd,mathrsfs,amsthm,slashed}

\usepackage[]{graphicx}
\usepackage{xcolor,dsfont}
\definecolor{darkblue}{rgb}{0.1,0.1,.7}
\usepackage[colorlinks, linkcolor=darkblue, citecolor=darkblue, urlcolor=darkblue, linktocpage]{hyperref} 
\usepackage[square, comma, compress,numbers]{natbib}
\usepackage{geometry}
\usepackage[margin=10pt,font=small,labelfont=bf]{caption}
\numberwithin{equation}{section}

\def\Ricci{\mca{R}}
\def\NN{\mathrm{N}}
\def\SS{\mathrm{S}}
\def\bm{\mathbf{m}}

\def\vol{\text{vol}}
\def\qaq{\quad \text{and} \quad}
\def\dag{\dagger}
\def\phs{{\phantom{*}}}
\def\phd{{\phantom{\dagger}}}
\def\wh{\widehat}

\def\wt{\widetilde}

\def\La{\Lambda}

\def\SO{\mathrm{SO}}
\def\Sd{\mathrm{S}_d}

\newcommand{\NO}[1]{{:\!#1\!:}}
\newcommand{\ud}[2]{^{#1}_{\phantom{#1}#2}}
\newcommand{\du}[2]{_{#1}^{\phantom{#1}#2}}

\newcommand{\Orange}{\color [rgb]{1,.5,0}}

\newcommand{\braket}[3]{\langle #1|#2|#3 \rangle}
\newcommand{\brakket}[2]{\langle #1|#2\rangle}
\newcommand{\ket}[1]{|#1\rangle}
\newcommand{\bra}[1]{\langle #1|}
\newcommand{\expec}[1]{\langle #1 \rangle}

\def\dps{\displaystyle}
\def\ldef{\mathrel{\mathop:}=}
\def\rdef{=\mathrel{\mathop:}}
\newcommand{\limu}[1]{\mathrel{\mathop{\sim}\limits_{\scriptstyle{#1}}}}

\def\fns{\footnotesize}
\newcommand{\reef}[1]{(\ref{#1})}
\def\beq{\begin{equation}} 
\def\eeq{\end{equation}}
\def\nn{\nonumber} 
\def\bsub{\begin{subequations}}
\def\esub{\end{subequations}}

\def\mbb{\mathbb}

\def\mca{\mathcal}

\def\mrm{\mathrm}
\def\msc{\mathscr}
\def\mtt{\mathtt}

\def\th{\tfrac{1}{2}}
\def\half{\frac{1}{2}}

\def\pd{\partial}
\def\a{\alpha}
\def\b{\beta}

\def\dd{\delta}
\def\ka{\kappa}
\def\la{\lambda}

\def\ze{\zeta}
\def\DD{\Delta}
\def\Oo{\mathcal{O}}

\def\l{\ell} 

\def\vareps{\varepsilon}
\def\bn{\mathbf{n}}
\def\unit{\mathds{1}} 
\newcommand{\threej}[6]{ \begin{pmatrix} #1 & #2 & #3 \\ #4 & #5 & #6 \end{pmatrix}}

\def\Red{\color [rgb]{0.9,0.1,0.1}}
\def\Green{\color [rgb]{0.3,0.5,0.2}}
\def\Blue{\color [rgb]{0.3,0.5,0.8}}

\begin{document}

\vspace*{-.6in} \thispagestyle{empty}
\begin{flushright}
 YITP-SB-18-32
\end{flushright}
\vspace{1cm} {\Large
\begin{center}
  {\bf RG flows on $S^d$ and Hamiltonian truncation}
\end{center}}
\vspace{1cm}
\begin{center}
{\bf Matthijs Hogervorst }\\[2cm] 
{
  Perimeter Institute for Theoretical Physics, Waterloo, ON, Canada \\
C.N.\@ Yang Institute for Theoretical Physics, Stony Brook University, USA
}
\\
\end{center}
\vspace{14mm}

\begin{abstract}
  We describe a nonperturbative method to compute the partition function and correlation functions for scalar QFTs set on the $d$-dimensional sphere $S^d$. The method relies on a Hamiltonian picture, where the theory is quantized on $S^{d-1}$ and states evolve in time by means of a time-dependent Hamiltonian. Crucially, the Hilbert space on $S^{d-1}$ is truncated to a finite set of states below a cutoff. Throughout this work we focus on the $\phi^2$ and $i\phi^3$ flows in three dimensions. In the first part of this paper we analyze the cutoff-dependence of various observables, computing both divergent and RG-improvement counterterms to be added to the action. Next we present nonperturbative results for the massive scalar on $S^3$, finding good agreement in the strong-coupling regime between numerical data and the $F$-coefficient of the free scalar CFT. We also check that the renormalized $i \phi^3$ theory on $S^3$ is nonperturbatively UV-finite. The scheme in question breaks the $\SO(d+1)$ spacetime symmetry group of $S^d$ down to $\SO(d)$, and in an example we study how the full symmetry is restored in the continuum limit. The relation between our method and earlier work by Al.\@ B.\@ Zamolodchikov involving a specific RG flow on $S^2$ is explained as well.
\end{abstract}
\vspace{12mm}

\newpage

{
\setlength{\parskip}{0.05in}
\tableofcontents
}

\setlength{\parskip}{0.05in}

\newpage

\section{Introduction}

This paper discusses $d$-dimensional Euclidean quantum field theories compactified on the sphere $S^d$. There are various reasons to study QFTs on this specific manifold. For one, the sphere is the only maximally symmetric compact manifold, so it provides a natural setting to study QFTs in finite volume. Second, certain non-local observables on the sphere are natural probes to study the QFT landscape. In two dimensions, the partition function of a QFT on the sphere is related to the $c$-coefficient, which obeys a famous monotonicity theorem~\cite{Zamolodchikov:1986gt} that severely constrains RG flows.\footnote{See also~\cite{Cardy:1988cwa} for an attempt to generalize the $c$-theorem to $d > 2$ dimensions.} More recently, it has been understood that the partition function on $S^3$ encodes a quantity $F$ that obeys a monotonicity theorem similar to $c$: specifically, an RG flow between two (unitary) CFTs can only exist if $F_\text{UV} > F_\text{IR}$~\cite{Casini:2012ei}.\footnote{Often, the same quantity $F$ is discussed in the context of entanglement entropy, see e.g.~\cite{Liu:2012eea,Klebanov:2012va,Casini:2015woa,Ben-Ami:2015zsa}.} Unfortunately, it is not known in general how to compute $F$ for general CFTs, and consequently the $F$-coefficient is only known for theories in certain corners of theory space: see \cite{Klebanov:2011gs,Anninos:2012ft,Fei:2014yja,Giombi:2014xxa,Fei:2015oha,Giombi:2015haa,Tarnopolsky:2016vvd} for results in free CFTs, CFTs at large $N$ or estimates that were obtained using the epsilon expansion, or~\cite{Jafferis:2010un,Jafferis:2011zi,Gulotta:2011si,Closset:2012vg} for theories with extended supersymmetry.

Let us make the relation between the partition function $Z_{S^3}$ and the $F$-coefficient explicit. Consider any renormalizable QFT, regulated by a local cutoff $\La$ and put on the sphere with radius $R$. At large $R$, the logarithm of the partition function reads
\beq
\label{eq:3dZ}
\ln Z_{S^3}(R) \, \limu{R \to \infty} \, B_1 (\La R)^3 + B_2  \La R - F + \ldots
\eeq
omitting terms that vanish as $R \to \infty$.\footnote{See Ref.~\cite{ZKnotes} for a pedagogical discussion.} The dimensionless coefficients $B_{1,2}$ are scheme-dependent and can be set to zero by adding local counterterms to the action (proportional to the cosmological constant and the Ricci scalar). Predicting $F$ for e.g.\@ the 3$d$ Ising CFT therefore requires computing $Z_{S^3}(R)$ for the $\phi^4$ theory with a fine-tuned mass term at large values of $R$, and subtracting the two terms with coefficients $B_{1,2}$. Since any RG flow initiated by a relevant operator is strongly coupled at large $R$, it follows that the above computation can only be performed using nonperturbative methods. %
 
It is therefore an interesting exercise to develop algorithms that allow for efficient QFT computations on $S^3$, or more generally on $S^d$. In principle, it is possible to perform Monte Carlo simulations on any latticized manifold with curvature. In practice, this is rather challenging: special care must be taken in order to guarantee that the QFT in question has the correct continuum limit as the lattice spacing is sent to zero. In recent years, Brower et al.~\cite{Brower:2016moq,Brower:2016vsl,Brower:2018szu} have made significant progress towards solving this problem and developing Monte Carlo algorithms for QFTs on $S^2$ and $\mbb{R} \times S^2$. For earlier work in the same vein, we point to Refs.~\cite{Brower:2012vg,Brower:2012mn,Brower:2014daa}.

In the present work a different approach is taken, based on the philosophy of Hamiltonian truncation. This is a variational approach to quantum field theory, see e.g.~\cite{truncRev} for a review of the subject and Refs.~\cite{Katz:2013qua,Hogervorst:2014rta,Katz:2014uoa,Rychkov:2014eea,Rychkov:2015vap,Elias-Miro:2015bqk,Katz:2016hxp,Balthazar:2016utu,Elias-Miro:2017tup,Elias-Miro:2017xxf,Whitsitt:2017ocl,Anand:2017yij,Rutter:2018aog,Fitzpatrick:2018ttk} for some recent examples. Concretely we consider $d$-dimensional scalar QFTs with polynomial interactions
\beq
\label{eq:Sact}
S[\phi] = S_0[\phi]  + \sum_n \frac{\la_n}{n!} \int_{S^d}\!\sqrt{g}d^dx\, \phi^n(x)
\eeq
where $S_0[\phi]$ is a Gaussian action (including a curvature coupling $\propto \Ricci \phi^2$). We proceed by canonically quantizing $\phi$ with respect to a foliation of $S^d$ that has $S^{d-1}$ timeslices.  This means that we treat the theory~\reef{eq:Sact} as an quantum system in $d-1$ dimensions with a time-dependent Hamiltonian. Instead of working with the full Fock space of the theory, we restrict to a large but finite number of states, characterized by a hard cutoff $\La$. After this truncation, it is possible to compute observables like the partition function nonperturbatively, and the exact answer is recovered by taking the continuum limit $\La \to \infty$. Our approach differs from most of the Hamiltonian truncation literature, where QFTs on manifolds of the form $\mbb{R} \times \mca{M}$ are considered. In such cases, ``solving'' the QFT amounts to estimating the spectrum of its Hamiltonian on $\mca{M}$ (or the transfer matrix, in the case of spin systems).

One interesting theory to study in this setting is the $\phi^2$ RG flow, i.e.\@ the massive boson on $S^d$. Since this theory is exactly solvable, it serves as a benchmark, where numerical data can be compared to analytic results. A second theory of interest is the $\phi^3$ interaction with imaginary coupling.\footnote{The $\phi^3$ theory with real coupling in $d$ dimensions does not exist nonperturbatively. The same theory with imaginary coupling has a $\mca{PT}$ symmetry~\cite{Bender:2012ea} that makes the theory well-defined.} Although the $i\phi^3$ theory is non-unitary, it provides a simple example of an interacting QFT in $d>2$. Many generic features of $d>2$ RG flows (notably UV divergences) are already present in this model. It is possible to flow from the free theory with an $i\phi^3$ interaction to the Yang-Lee CFT, but reaching the critical point requires a UV finetuning, and this is left for future work. 

This paper is organized as follows. In Sec.~\ref{sec:can} we define a specific foliation of $S^d$ that will be used to quantize the scalar field $\phi$ on $S^d$. In Sec.~\ref{sec:inter}, an explicit procedure to compute the partition function and certain correlation functions is described, and the method in question is compared to earlier work involving minimal model flows on $S^2$. All observables measured in this scheme depend on the cutoff $\La$; in Sec.~\ref{sec:cutoff} this cutoff-dependence will be studied in perturbation theory, and various counterterms that are generated are computed. In Sec.~\ref{sec:numerics}, the method is tested numerically, for both the $\phi^2$ and $i\phi^3$ theories on $S^3$. Finally, the breaking of $\SO(d+1)$ at the cutoff scale is examined in Sec.~\ref{sec:onept}.

\section{Gaussian theory on $S^d$}
\label{sec:can}

In this section we will discuss the canonical quantization of a non-interacting scalar theory on $S^d$. We start by defining a foliation of $S^d$, and proceed by explicitly developing the necessary canonical quantization.

\subsection{Geometry}
\label{sec:geom}

\begin{figure}[htbp]
\begin{center}
\includegraphics[scale=0.88]{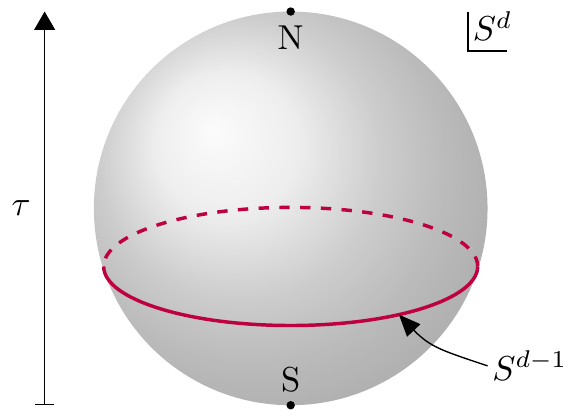}
\end{center}
\caption{{ Foliation of $S^d$ where every leaf is a copy of $S^{d-1}$. Euclidean time $\tau \in \mbb{R}$ runs upwards from the South to the North pole, with $\tau = 0$ describing the equator.}}
\label{fig:sphere}
\end{figure}

The foliation of $S^d$ used in this paper is defined as follows. We take the leafs (or timeslices) to be parallel copies of $S^{d-1}$, parametrized by a spherical coordinate $\bn$. The remaining Euclidean time coordinate is denoted as $\tau \in \mbb{R}$. Explicitly, we define $S^d \subset \mbb{R}^{d+1}$ by means of the following parametrization:
\beq
\label{eq:param}
X^\mu(\tau,\bn) = \frac{R}{\cosh \tau} \left(\sinh \tau, \bn \right) \in \mbb{R}^{d+1},
\quad
\dd_{\mu \nu}X^\mu X^\nu = R^2.
\eeq
The points $\tau = \pm \infty$ correspond to the North and South poles of $S^d$, whereas $\tau = 0$ parametrizes the equator. A sketch of this foliation is shown in Fig.~\ref{fig:sphere}. 
In the coordinates of~\reef{eq:param} the induced metric on $S^d$ reads
\beq
\label{eq:metric}
ds^2 = \left(\frac{R}{\cosh \tau}\right)^2 \left[d\tau^2 + d\Omega(\bn)^2 \right] 
\eeq
where $d\Omega(\bn)^2$ is the standard metric of a unit-radius $S^{d-1}$. In particular, Eq.~\reef{eq:metric} exhibits that the sphere is topologically equivalent to the $d$-dimensional cylinder $\mbb{R} \times S^{d-1}$.

Let us proceed by stating some technical results that will be necessary in the rest of this paper. First, we recall that the scalar curvature of the sphere is constant and given by $\Ricci = d(d-1)/R^2$.  In the $(\tau,\bn)$ coordinates, the Laplacian on $S^d$ is given by
\beq
\label{eq:lapl}
\nabla^2 = \left(\frac{\cosh \tau}{R} \right)^2 \left[ \frac{\pd^2}{\pd \tau^2}  -(d-2) \tanh \tau \frac{\pd}{\pd \tau} - \vec{L}^2 \right]\!.
\eeq
Here $\vec{L}^2$ is the positive Laplacian on $S^{d-1}$, having eigenvalues $\l(\l+d-2)$.\footnote{See~\cite{harm} for a comprehensive discussion of this operator.} The multiplicity of the $\l$-th eigenvalue is
\beq
n_\l^d =  \frac{(2\l+d-2)(\l+d-3)!}{\l!(d-2)!},
\quad
\l = 0,1,2,\ldots
\eeq
and we denote the corresponding eigenfunctions --- the (hyper)spherical harmonics --- by $Y_{\l j}(\bn)$. We refer to Appendix~\ref{sec:harmonics} for the conventions used in this paper and some useful identities..

The isometry group of the $d$-sphere is $\SO(d+1)$, contrasting with the Poincar\'{e} group (or its Euclidean counterpart) in flat space. A subgroup $\SO(d) \subset \SO(d+1)$ acts only on the spatial coordinate $\bn$. The remaining $d$ generators mix space and time, and are the counterpart of the  generators $P_\mu$ of translations in flat space. We stress that the generator of time translations $\pd/\pd \tau$ is \emph{not} a Killing vector, which means that there is no notion of (conserved) energy.  Physically, the isometries of $S^d$ impose nontrivial constraints on correlation functions~\cite{Osborn:1999az}. If we consider a correlator of $n$ scalar operators $\Oo_i$, these constraints take the form of Ward identities:
\beq
\label{eq:Wardid}
\sum_{i=1}^n L_a^{(i)} \cdot \expec{\Oo_1(\tau_1,\bn_1) \dotsm \Oo_n(\tau_n,\bn_n)} = 0
\eeq
where  $\dps{L_a^{(i)}}$ is a first-order differential operator acting on the insertion $\Oo_i(\tau_i,\bn_i)$. There is one such equation for every generator $L_a$ of $\SO(d+1)$. For $n=1$, the constraint~\reef{eq:Wardid} implies that every scalar one-point function $\expec{\Oo(\tau,\bn)}$ is constant on $S^d$. Two-point functions may only depend on the \emph{chordal distance} $\msc{S}$ between two points:
\beq
\label{eq:2pt}
\expec{\Oo_1(\tau_1,\bn_1)\Oo_2(\tau_2,\bn_2)} = f(\msc{S})
\eeq
where $f$ is an arbitrary function  and
\beq
\label{eq:chord}
\msc{S} \ldef \frac{1}{(2R)^2}\|X^\mu(\tau_1,\bn_1) - X^\mu(\tau_2,\bn_2)\|^2 =  \frac{\cosh(\tau_1 - \tau_2) - \bn_1\cdot {\bf n}_2}{2\cosh \tau_1 \cosh \tau_2}\,.
\eeq
By construction, $\msc{S}$ is invariant under $\SO(d+1)$ transformations and bounded: $0 \leq \msc{S} \leq 1$. More complicated constraints apply to correlators of $n > 2$ operators and spinning correlators, but such Ward identities will not play a role in the present work.

\subsection{Canonical quantization}
\label{sec:boson}

Let us proceed by setting up the canonical quantization of a scalar field in $d \geq 2$ dimensions with respect to the foliation described in Sec.~\ref{sec:geom}. As a starting point, consider the conformally coupled scalar boson, described by the following action:
\beq
\label{eq:L0}
S_0[\phi] = \half \int_{S^d}\!\sqrt{g}d^dx \left[g^{\mu \nu}\pd_\mu \phi \pd_\nu \phi +    \xi_c  \Ricci \, \phi^2 +  m^2 \phi^2\right],
\quad
\xi_c = \frac{d-2}{4(d-1)}.
\eeq
By construction the above action is conformally invariant if $m^2 = 0$, although we will keep the bare mass $m$ general for now. Working in arbitrary coordinates $x^\mu$, the $\expec{\phi(x_1) \phi(x_2)}$ two-point function is the unique solution to the Klein-Gordon equation
\beq
\label{eq:KG}
\left[- \nabla_{x_1}^2 + \xi_c \Ricci +  m^2  \right]\!\expec{\phi(x_1)\phi(x_2)} = \frac{\dd^{(d)}(x_1,x_2)}{\sqrt{g(x_1)}}
\eeq
that is regular on $S^d$, away from the coincident limit $x_1 \to x_2$. The solution can be written in compact form as 
\begin{multline}
\label{eq:prop1}
\expec{\phi(\tau_1,\bn_1)\phi(\tau_2,\bn_2)} =  \frac{1}{R^{d-2}} \frac{\Sd}{2(2\pi)^d}  \Gamma(\th(d-1) + \zeta) \Gamma(\th(d-1) - \zeta) \\
\times  \frac{1}{\msc{S}^{\half d-1}} \, {}_2F_1(\th + \zeta, \th-\zeta; \th d; 1-\msc{S}),
\quad\;
\zeta \ldef \sqrt{\frac{1}{4}-m^2R^2}.
\end{multline}
Here $\Sd \ldef \vol(S^{d-1}) = 2\pi^{\half d}/\Gamma(\th d)$ and $\msc{S}$ is the chordal distance introduced in Eq.~\reef{eq:chord}. Notice that depending on the value of $m$, $\ze$ is either real or purely imaginary; nonetheless, the Green's function~\reef{eq:prop1} is manifestly real, since it is invariant under $\zeta \mapsto -\zeta$.

At the same time, the action~\reef{eq:L0} can be quantized using a well-known recipe~\cite{Birrell:1982ix}. We will leave the details of this computation to the reader and merely state the results. The mode decomposition of $\phi$ is given by
\beq
\label{eq:phidec}
\phi(\tau,\bn) = \frac{1}{R^{\half d-1}}\sum_{\l=0}^\infty \sum_{j=1}^{n_\l^d} a^\phd_{\l j} \, K_\l^\phs\!(\tau) Y_{\l j}^\phs(\bn) + a^\dagger_{\l j} \, K_\l^\phs\!(-\tau) Y_{\l j}^*(\bn)
\eeq
where the $K_\l(\tau)$ are certain mode functions, to wit:
\begin{multline}
\label{eq:Kl}
K_\l(\tau) = \frac{ \sqrt{\th \Gamma(\l+\th(d-1) +\zeta)\Gamma(\l+\th(d-1) - \zeta)}}{\Gamma(\l+\half d)}  \\
\times \, (\cosh \tau)^{\half d-1} \, e^{-(\l+\half d-1)\tau}  \, {}_2F_1\!\left[{{\half + \zeta,\, \half-\zeta}~\atop~{\l+\half d}}\,  \Bigg| \, \frac{1-\tanh\tau}{2} \right]\!.
\end{multline}
The creation and annihilation operators in~\reef{eq:phidec} obey canonical commutation relations:
\beq
\label{eq:CCR}
[a^\phd_{\l j},\, a^\dagger_{\l' j'}] = \dd_{\l \l'} \dd_{j j'},
\quad
[a_{\l j},\, a_{\l' j'}] = [a^\dag_{\l j},\, a^\dag_{\l' j'}] = 0.
\eeq
Using Eq.~\reef{eq:lapl}, it can easily be checked that~\reef{eq:phidec} is a solution to the Klein-Gordon equation. Remark that the field $\phi$ defined in~\reef{eq:phidec} is Hermitian in the Euclidean sense, meaning that $\phi$ obeys
\beq
\phi(\tau,\bn)^\dagger = \phi(-\tau,\bn).
\eeq
The mode decomposition~\reef{eq:phidec} can be checked  by computing the time-ordered two-point function of $\phi$ inside the Fock space vacuum $\ket{\emptyset}$ and comparing the result to~\reef{eq:prop1}. This yields
\begin{multline}
\label{eq:prop2}
\braket{\emptyset}{\mrm{T}\phi(\tau_1,\bn_1)\phi(\tau_2,\bn_2)}{\emptyset} = \frac{1}{R^{d-2}} \sum_{\l=0}^\infty \frac{n_{\l}^d}{\Sd} \, C_\l^d(\bn_1 \cdot \bn_2) \Big[ \Theta(\tau_1 > \tau_2) K_\l(\tau_1)K_\l(-\tau_2) \\
  + \Theta(\tau_2 > \tau_1)K_\l(\tau_2)K_\l(-\tau_1) \Big]
\end{multline}
where the $C_\l^d(z)$ denote $d$-dimensional Gegenbauer polynomials, as defined in Appendix~\ref{sec:harmonics}. Both expressions agree, as can for instance be checked numerically.

It is straightforward to define composite operators, provided that one normal-orders all creation and annihilation operators in the relevant mode decompositions. For instance, the operator $\NO{\phi^n}$ is defined as
\beq
\label{eq:noop}
\NO{\phi^n(\tau,\bn)} = \sum_{k=0}^n \binom{n}{k} \phi_+^k(\tau,\bn) \phi_{-}^{n-k}(\tau,\bn) \neq [\phi(\tau,\bn)]^n
\eeq
where $\phi_+$ (resp.\@ $\phi_{-}$) denotes the part of $\phi$ containing creation (annihilation) operators. Since we will always normal-order composite operators, we will omit the $\NO{\Oo}$ notation from now on and write $\Oo$ instead. Time-ordered correlation functions of composite operators can be computed using the canonical commutation relations, similar to~\reef{eq:prop2}. 

\section{Turning on interactions}\label{sec:inter}

Next, let us turn our attention to interacting scalar theories. We start by expressing the partition function of an interacting QFT in terms of a time-evolution operator $U$. We then discuss a method to systematically approximate this operator in detail.

\subsection{Hamiltonian picture}
\label{sec:hamont}

Let us consider a general interacting QFT, e.g.\@ having an action of the form~\reef{eq:Sact}. More generally, we consider an exactly solvable theory $S_0$, either a Gaussian theory or a CFT, that is perturbed by a local operator $\Oo$:
\beq
\label{eq:int}
S = S_0 + \dd S,
\quad
\dd S = \la \int_{S^d}\!\sqrt{g}d^dx \, \Oo(x).
\eeq
We assume that $\Oo$ is relevant, meaning that we can assign a scaling dimension $\DD < d$ to $\Oo$, and the coupling $\la$ has mass dimension $[\la] = d-\DD$. Working in the $(\tau,\bn)$ coordinates from Sec.~\ref{sec:geom}, we can rewrite the interaction term~\reef{eq:int} as
\beq
  \dd S   = \bar\la \int_\mbb{R} \frac{d\tau}{(\cosh \tau)^d} \, \wh{\Oo}(\tau) \quad \text{where} \quad \wh{\Oo}(\tau) \ldef R^{\DD} \int_{S^{d-1}}\!d\bn\, \Oo(\tau,\bn) \label{eq:Ohat}
\eeq
and $\bar\la$ denotes the dimensionless coupling $\bar{\la} \equiv \la R^{d-\DD}$. In the path integral picture, the partition function of the interacting theory is given by
\beq
\label{eq:Zla}
Z(\la) = \int\![d\phi]  e^{-S_0 - \dd S}\,.
\eeq
In principle, Eq.~\reef{eq:Zla} can be expressed as an infinite sum of integrated correlation functions in the $\la = 0$ theory. Explicitly:
\beq
\label{eq:Zexp1}
Z(\la) = \sum_{n=0}^\infty \frac{(-\bar{\la})^n}{n!} \prod_{i=1}^n \int_{-\infty}^\infty \frac{d\tau_i}{(\cosh \tau_i)^d} \, \expec{\wh{\Oo}(\tau_1) \dotsm \wh{\Oo}(\tau_n)}_0 \, Z(0).
\eeq
The notation $\expec{\mca{X}}_0$ denotes correlators that are measured in the free theory, i.e.
\beq
\expec{\mca{X}}_0 = \frac{1}{Z(0)} \int\![d\phi]  \mca{X} e^{-S_0}
\eeq
where $\mca{X}$ is any string of local operators. As is customary, we can work with time-ordered correlators instead:
\beq
\label{eq:Zexp2}
\frac{Z(\la)}{Z(0)} = \sum_{n=0}^\infty (-\bar{\la})^n \prod_{i=1}^n \int_{\tau_i < \tau_{i+1}}\! \frac{d\tau_i}{(\cosh \tau_i)^d} \, \expec{\wh{\Oo}(\tau_n) \dotsm \wh{\Oo}(\tau_1)}_0
\eeq
at the expense of omitting the $1/n!$ factor appearing in Eq.~\reef{eq:Zexp1}. Expressions like~\reef{eq:Zexp2} are ubiquitous in the interaction picture of quantum mechanics. To proceed, we interpret the correlation functions in~\reef{eq:Zexp2} as matrix elements, cf.\@ the discussion from the previous section:
\beq
\expec{\wh{\Oo}(\tau_n) \dotsm \wh{\Oo}(\tau_1)}_0 = \braket{\emptyset}{\mrm{T} \wh{\Oo}(\tau_1) \dotsm \wh{\Oo}(\tau_n)}{\emptyset}.
\eeq
Then the RHS of~\reef{eq:Zexp2} can then be interpreted as a matrix element of a time-evolution operator $U$, to wit:
\beq
\label{eq:Udef1}
\boxed{
\frac{Z(\la)}{Z(0)} = \braket{\emptyset}{U(\infty,-\infty)}{\emptyset},
\quad
U(\tau_f,\tau_i) = \mrm{T} \exp\!\left[ - \bar{\la} \int_{\tau_i}^{\tau_f}\! \frac{d\tau}{(\cosh \tau)^d}\,  \wh{\Oo}(\tau)\right]\!.
}
\eeq
In quantum mechanics, $U$ is often referred to as a \emph{Dyson operator}.

At a first glance, Eq.~\reef{eq:Udef1} seems to be nothing but a rewriting of the perturbative expansion~\reef{eq:Zexp1}. In practice, the above formula provides a powerful way to evaluate $Z(\la)$ numerically. The basic idea is to divide the sphere into many timeslices. Concretely, given any two times $\tau_i < \tau_f$ we can write the Dyson operator on the interval $[\tau_i,\tau_f]$ as
\beq
\label{eq:trotter1}
U(\tau_f,\tau_i) = \prod_{k=0}^{T-1} U(\tau_k + \th \dd \tau,\tau_k - \th \dd \tau),
\quad
\tau_k = \tau_i + (k+\th)\dd\tau,
\eeq
where $\dd\tau \equiv (\tau_f - \tau_i)/T$. The product in~\reef{eq:trotter1} runs from right to left, i.e.\@ $k=0$ (resp.\@ $k=T-1$) is the rightmost (leftmost) factor. Note that the formula~\reef{eq:trotter1} is exact, regardless of the number of timeslices $T$.

A second important idea is that for sufficiently large $T$, every individual factor in~\reef{eq:trotter1} can be replaced by a first-order approximation in $\dd \tau$. Crucially, this approximation induces an error of order $1/T$, hence the operator $U(\tau_f,\tau_i)$ is recovered in the limit $T \to \infty$. Quantitatively, this statement can be phrased as the following identity
\beq
\label{eq:trotter}
U(\tau_f,\tau_i) =  \prod_{k=0}^{T-1} \left[\unit - \frac{\bar\la \dd\tau}{(\cosh \tau_k)^d} \, \wh{\Oo}(\tau_k)\right] + O(1/T)
\eeq
which is known as a \emph{Trotter} or \emph{Suzuki-Trotter formula}. A simple derivation of this result will be given in Sec.~\ref{sec:discerror}. In the next section, we will explain how formula~\reef{eq:trotter} can be used in practice to compute $Z(\la)$.

\subsection{Numerical setup and truncation}
\label{sec:truncgen}

The Trotter formula~\reef{eq:trotter} provides a starting point for a numerical evaluation of matrix elements of $U$. Every individual factor in this formula involves an infinite-dimensional operator $\wh{\Oo}(\tau)$. In the scalar theories studied in this paper, these operators act on the Fock space $\msc{F}$ of creation operators $a^\dag_{\l j}$. Between every two factors in~\reef{eq:trotter}, a resolution of the identity can be inserted: if
\beq
u(\tau) \ldef \unit - \frac{\bar\la \dd\tau}{(\cosh \tau)^d} \, \wh{\Oo}(\tau)
\eeq
then
\beq
\label{eq:resid}
 u(\tau_{k+1}) u(\tau_k)  = \sum_{\ket{\psi} \in \msc{F}} u(\tau_{k+1}) \ket{\psi} \bra{\psi} u(\tau_k)
\quad
\eeq
assuming that the states $\ket{\psi}$ form an orthonormal basis of $\msc{F}$. By using~\reef{eq:resid} repeatedly, the Trotter formula reduces the computation of the partition function to an expression involving an infinite number of matrix elements $\braket{\psi_i}{\wh{\Oo}(\tau)}{\psi_j}$. In the present work, we instead truncate to a finite subset of states $\msc{F}(\La) \subset \msc{F}$ depending on a UV cutoff $\La$ to be defined later. Instead of the resolution of the identity in~\reef{eq:resid}, we therefore use the approximation
\beq
\label{eq:resalt}
u(\tau_{k+1}) u(\tau_k) \, \longrightarrow  \, u(\tau_{k+1}) P_\La  u(\tau_{k}),
\quad
P_\La = \sum_{\ket{\psi} \in \msc{F}(\La)} \ket{\psi} \bra{\psi}.
\eeq
The subspace $\msc{F}(\La)$ consists of all states with a certain ``energy'' $E(\ket{\psi}) \leq \La$. The term energy is a misnomer, since there is no notion of energy on $S^d$, but with a slight abuse of language we will use this term from now on. To define $E$, it is necessary to organize the Fock space into basis states of the following form
\beq
\label{eq:basisState}
\ket{\psi} = \sum_{\{j_i\}} T^{j_1 \dotsm j_n}_\psi  a_{\l_1 j_1}^\dag \!\!\dotsm a^\dag_{\l_n j_n} \ket{\emptyset}
\eeq
where $T_\psi$ is a tensor with labels in the (symmetrized) $\l_1 \otimes \ldots \otimes \l_n$ representation of $\SO(d)$. For a state of the form~\reef{eq:basisState}, the energy $E(\ket{\psi})$ is defined as follows:\footnote{For $d=2$, the truncated Hilbert space $\msc{F}(\La)$ is only finite if $m^2 > 0$. For $d>2$, $\msc{F}(\La)$ is always finite.}
\beq
E(\ket{\psi}) = \sum_{i=1}^n \vareps(\l_i),
\quad
\vareps(\l) = \sqrt{\l(\l+d-2)/R^2 + m^2 + d(d-2)/(4R^2)}.
\eeq
Roughly speaking, $\vareps(\l)$ measures the single-particle energy of a spin-$\l$ mode, measured at the equator. Indeed, the operator ${\mrm{D} = -\nabla^2 + m^2 + \xi_c \Ricci}$ acts on spherical harmonics as
\[
\mrm{D} \cdot Y_{\l j}(\bn) = \left[ (\cosh \tau)^2 \, \l(\l+d-2)/R^2 + m^2 + \xi_c \Ricci \right] \! Y_{\l j}(\bn).
\]
The quantity inside brackets equals $\vareps(\l)^2$ at $\tau = 0$ and is strictly larger for all other $\tau$. Clearly, there are many other possible ways to truncate the Fock space, and perhaps a different truncation procedure is more efficient. We leave this question open for future work.

Summarizing, there are two different cutoffs used in this work, $\La$ and $T$. The former has a simple physical interpretation: it is a hard cutoff in energy, which means that correlation functions can only be resolved up to timescales $\DD \tau \gtrsim 1/\La R$. On the other hand, the number of timeslices $T$ is an unphysical regulator that only controls the discretization error in the Trotter formula~\reef{eq:trotter}.

So far, we have glanced over one important detail. In Eqs.~\reef{eq:trotter1} and~\reef{eq:trotter}, we chose the sampling times $\{\tau_k\}$ uniformly over the interval $[\tau_i,\tau_f]$. Such a uniform sampling is not feasible in practice, since the endpoints $\tau_f,\tau_i$ must be sent to $\pm \infty$ when the partition function is measured. Rather, we introduce a new time coordinate $z = z(\tau) \in [0,1]$ as follows
\beq
\label{eq:zdef}
z(\tau) \ldef \frac{\mrm{S}_d}{\mrm{S}_{d+1}} \int_{-\infty}^\tau \frac{d\tau'}{(\cosh \tau')^d}
\eeq
and from now on we will sample uniformly in $z$. To be completely explicit, this means that Trotter formulas like Eq.~\reef{eq:trotter} will be evaluated at times $z_k = (k+\th)/T$ for ${k=0,1,\ldots,T-1}$. In order not to overload the notation, we will mostly use the notation $\tau$ instead of $z$ throughout this paper.

\subsection{Discretization error}
\label{sec:discerror}

In this section, we will present a heuristic derivation of the Trotter formula~\reef{eq:trotter}. Keeping the UV cutoff $\La$ fixed, we will argue that the discretization error in~\reef{eq:trotter} vanishes as $1/T$ in the limit $T \to \infty$, although the Trotter formula applies to a more general class of operators~\cite{simonb}. This derivation is unimportant for the rest of this work, and can be skipped on a first reading. 

After truncating, the operators $U(\tau',\tau)$ and $\wh{\Oo}(\tau)$ are matrices that act on the finite-dimensional vector space $\msc{F}(\La)$. Let $\lVert \cdot \rVert$ denote the $L_2$ norm on the space of such matrices, $\lVert \Oo \rVert^2 = \text{tr}\,  {}^t\Oo \Oo$. On a sufficiently small interval of size $\dd \tau$, the error made using the Trotter approximation can be bounded as follows:
\begin{multline}
  \label{eq:error}
\left\lVert \, U(\tau+\th \dd \tau,\tau-\th \dd\tau) - \left(\unit - \frac{\bar\la \dd\tau}{(\cosh \tau)^d} \, \wh{\Oo}(\tau)  \right)
\right\rVert\\
\leq \; \half \frac{\bar\la^2 \dd \tau^2}{(\cosh \tau)^{2d}}\,  \lVert \wh{\Oo}(\tau) \rVert^2 \, + \, \bar\la \dd \tau^2 \left\lVert \frac{d}{d\tau} \frac{\wh{\Oo}(\tau)}{(\cosh \tau)^d} \right\rVert + O(\dd \tau^3).
\end{multline}
The first term on the RHS of~\reef{eq:error} originates from replacing the function $\exp(\cdot)$ appearing in the evolution operator $U$ by its linearization. The second term arises because $\Oo$ is evaluated at time $\tau$, instead of integrating over the interval $[\tau - \th \dd \tau, \tau + \th \dd\tau]$. A precise estimate of either of these terms is not important. However, for definiteness, we find that for operators $\Oo = \phi^n$ in $d=3$, there exists a rough bound of the following form:
\beq
\label{eq:expbbb}
d=3:
\quad
\lVert \wh{\phi^n}(\tau) \rVert \lesssim b (\cosh \tau)^{n/2} \, \exp(c \La^{2/3}),
\quad
b,c = O(1).
\eeq
The $\exp(c \La^{2/3})$ growth reflects the number of states below the cutoff (i.e.\@ the size of the matrix $\wh{\Oo}(\tau)$).  Likewise, a precise bound on the second term will not be important, but a rough estimate gives
\beq
\label{eq:varerror}
\left\lVert \frac{d}{d\tau} \wh{\Oo}(\tau) \right\rVert \lesssim \La    \lVert \wh{\Oo}(\tau)\rVert
\eeq
setting $R=1$ from now on. Near $\tau = \pm \infty$, there will be finite corrections to the RHS of~\reef{eq:varerror}.
Bring everything together, we find that
\begin{align}
  \label{eq:mds}
  \text{LHS of Eq.~\reef{eq:error}} \, &\lesssim \,  \bar\la^2 \dd \tau^2 \exp(2c \La^{2/3})  + \bar\la \dd \tau^2  \exp(c \La^{2/3}+ \ln \La) + O(\dd \tau^3)\\
  &= f  \dd \tau^2 + O(\dd \tau^3) \nn
\end{align}
where $f = f(\bar\la,\La)$ depends on the coupling $\la$ and the UV cutoff $\La$, but not on the size $\dd \tau$ of the interval. In~\reef{eq:mds} we have set $\tau \simeq 0$ for simplicity and omitted various constants. Since there are $T$ timeslices in total, the total error made by discretizing is of the order of $T \cdot \dd \tau^2 \sim 1/T$, as we set out to prove.

We have been somewhat cavalier about the $\tau$-dependence of the operator $\wh{\Oo}(\tau)$. This is justified away from the poles at $\tau = \pm \infty$, since the norm of both the operator $\wh{\Oo}(\tau)$ and its $\tau$-derivative can be uniformly bounded on any compact interval. However, the Trotter formula can break down when wave functions blow up near the poles. More quantitatively, if $\ket{\Psi}$ is a given in-state, it may happen that the time-evolved state
\beq
 U(\tau,-\infty) \ket{\Psi} = \ket{\Psi} - \bar\la \int_{-\infty}^\tau\!\frac{d\tau'}{(\cosh \tau')^d} \, \wh{\Oo}(\tau) \ket{\Psi} + \ldots
\eeq
fails to be normalizable, reflecting a physical UV divergence. For the matrix elements studied in this work, such divergences do not occur. 

\subsection{Other observables}

The above derivation extends to correlation functions in a straightforward way. Suppose that we want to compute the one-point function of a scalar operator $\Oo'$:
\beq
\expec{\Oo'(\tau,\bn)}_\text{disc} = \frac{1}{Z(0)} \int\![d\phi] \Oo'(\tau,\bn) \, e^{-S_0 - \dd S}.
\eeq
We use the notation $\Oo'$ to make clear that this operator is unrelated to the operator $\Oo$ appearing in the action. 
Following the same steps as above, we find that
\beq
\label{eq:Adisc}
\expec{\Oo'(\tau,\bn)}_\text{disc} = \braket{\emptyset}{U(\infty,\tau)\Oo'(\tau,\bn)U(\tau,-\infty)}{\emptyset}.
\eeq
In this normalization $\expec{\unit}_\text{disc} = Z(\la)/Z(0) \neq 1$; it is therefore more natural to measure connected correlation functions:
\beq
\label{eq:Aconn}
\expec{\Oo'(\tau,\bn)}_\text{conn} = \frac{Z(0)}{Z(\la)}\, \expec{\Oo'(\tau,\bn)}_\text{disc} =  \frac{\braket{\emptyset}{U(\infty,\tau)\Oo'(\tau,\bn)U(\tau,-\infty)}{\emptyset}}{\braket{\emptyset}{U(\infty,-\infty)}{\emptyset}}.
\eeq
In Eqs.~\reef{eq:Adisc} and~\reef{eq:Aconn} we can take the limit $\tau \to \mp \infty$, which corresponds to inserting $\Oo'$ at the South (resp.\@ North) pole:
\beq
\label{eq:Apoles}
\lim_{\tau \to -\infty} \braket{\emptyset}{U(\infty,\tau)\Oo'(\tau,\bn)U(\tau,-\infty)}{\emptyset} = \braket{\emptyset}{U(\infty,-\infty)}{\Oo'},
\quad
\ket{\Oo'} \ldef \lim_{\tau \to -\infty} \Oo'(\tau,\bn) \ket{\emptyset}.
\eeq
Provided that $\Oo'$ is normal-ordered, it is easy to show that the in-state $\ket{\Oo'}$ has finite norm and is independent of $\bn$. The limit $\tau \to \infty$ can be treated similarly.

We can similarly consider higher-point functions. In general, measuring higher-point functions requires intermediate states that are not $\SO(d)$ scalars. This is technically rather challenging in the framework described in this paper --- see Sec.~\ref{sec:implementation}. However, we can measure antipodal two-point functions of $\phi$. With an antipodal two-point function, we mean a correlator such as~\reef{eq:2pt} measured at $\msc{S} = 1$.  One way to compute such correlators is to insert the operators $\phi$ at both poles:
\beq
\label{eq:NS1}
 \expec{\phi(\NN)\phi(\SS)}_\text{disc} =   \braket{\phi(\mrm{N})}{U(\infty,-\infty)}{\phi(\mrm{S})}
\eeq
where the in- and out-states are defined as in Eq.~\reef{eq:Apoles}. The connected two-point function can be constructed as in~\reef{eq:Aconn}. 

\subsection{Comparison to Ref.~\cite{Zamolodchikov:2001dz}}

The approach we have described above is closely related to earlier work by Al.\@ B.\@ Zamolodchikov~\cite{Zamolodchikov:2001dz} (which appeared in a slightly different form as~\cite{zamolBook}).  In Ref.~\cite{Zamolodchikov:2001dz}, the author considered a $\DD = -2/5$ perturbation of the two-dimensional Yang-Lee CFT on $S^2$; the interacting partition function in that theory was computed by numerically integrating a time-dependent Schr\"{o}dinger equation. In a subsequent paper, Grinza and Magnoli~\cite{Grinza:2003ji} studied a $\DD = 1/8$ perturbation of the $2d$ Ising model using the same technique. In this section, we will spell out the connection between these $2d$ CFT papers and our method. What follows is independent from the rest of this paper and can be skipped on a first reading.

In the discussion that follows, two simple facts about CFTs will be important. First, we use that any conformally invariant theory can be mapped from $S^d$ to the cylinder $\mbb{R} \times S^{d-1}$ by a Weyl transformation (see e.g.~\cite{Rychkov:2016iqz,Simmons-Duffin:2016gjk}). Concretely, a (scalar) local operator $\Oo$ on $S^d$ maps to its counterpart $\Oo_\text{cyl}$ on the cylinder as follows:
\beq
\label{eq:cft1}
\Oo(\tau,\bn) = (\cosh \tau)^\DD\, \Oo_\text{cyl}(\tau,\bn)
\eeq
where $\DD$ is the scaling dimension of $\Oo$. Second, we recall that time translations on the cylinder are generated by the CFT dilatation operator $D$:
\beq
\label{eq:cft2}
\Oo_\text{cyl}(\tau,\bn) = e^{D \tau} \Oo_\text{cyl}(0,\bn) e^{-D \tau}.
\eeq
Bringing Eqs.~\reef{eq:cft1} and~\reef{eq:cft2} together, we can write every individual factor in Eq.~\reef{eq:trotter} as
\beq
\unit - \frac{\bar\la \dd \tau}{(\cosh \tau_k)^d} \, \wh{\Oo}(\tau_k) = e^{D \tau_k}\! \left[ \unit - \frac{\bar\la \dd \tau}{(\cosh \tau_k)^{d-\DD}} \, \wh{\Oo}_\text{cyl}(0) \right]\! e^{-D \tau_k}.
\eeq
Consequently, we obtain
\beq
\label{eq:CFTtrotter}
\bra{\Omega}U(\tau_f,\tau_i)\ket{\Omega}   =  \lim_{T \to \infty} \, \bra{\Omega}  \prod_{k=0}^{T-1} \left\{ \unit - \dd \tau\! \left[D + \frac{\bar\la}{(\cosh \tau_k)^{d-\DD}} \, \wh{\Oo}_\text{cyl}(0) \right] \right\} \ket{\Omega}
 \eeq
 where $\ket{\Omega}$ is the radial quantization ground state, obeying $D \ket{\Omega} = 0$. In obtaining this formula, we use that
\[
e^{-D \tau_k} e^{D \tau_{k-1}} = \unit - \dd \tau D + O(\dd \tau^2).
\]
Finally, if we define
\beq
\ket{\psi(\tau)} \ldef U(\tau,-\infty)\ket{\Omega}
\eeq
then Eq.~\reef{eq:CFTtrotter} implies that $\ket{\psi(\tau)}$ satisfies the following Schr\"{o}dinger equation:
\beq
-\!\frac{d}{d\tau} \ket{\psi(\tau)} = \left[D + \frac{\bar\la}{(\cosh \tau)^{d-\DD}} \, \wh{\Oo}_\text{cyl}(0) \right]\!\ket{\psi(\tau)}
\eeq
and the interacting partition function is recovered as the limit $Z(\la)/Z_\text{CFT} = \lim_{\tau \to \infty} \brakket{\Omega}{\psi(\tau)}$. After setting $d=2$ this is essentially formula Eq.~(5.9) from Ref.~\cite{Zamolodchikov:2001dz}.\footnote{In that work the notation $\DD$ refers to the eigenvalue of the $L_0$ generator of the Virasoro algebra, hence ${\DD_\text{here} = 2\DD_\text{there}}$.} An important feature of~\reef{eq:CFTtrotter} is that the time coordinate $\tau$ only features via the kinematical factor $1/(\cosh \tau)^{d-\DD}$: the perturbation $\wh{\Oo}_\text{cyl}(0)$ is a time-independent matrix. Moreover, the matrix elements of $\wh{\Oo}_\text{cyl}$ can be expressed as OPE coefficients of the UV CFT in question.

\section{Cutoff effects}
\label{sec:cutoff}

So far, we proposed a method to compute the partition function and correlation functions of scalar QFTs on $S^d$. Crucially, we truncated the Fock space to a finite-dimensional subspace characterized by a hard cutoff $\La$. Since generic QFTs have UV divergences, additional $\La$-dependent counterterms  must be added to the action to regulate the theory in the UV.\footnote{Since $S^d$ is a compact manifold, there are no IR divergences.} Moreover, in numerical computations it is not possible to send $\La \to \infty$, and working at finite $\La$ induces a systematic error (similar to working at finite lattice spacing). Adding counterterms that vanish as $\La \to \infty$ can reduce such systematic errors, cf.\@ Symanzik's RG-improved lattice actions~\cite{Symanzik:1983dc}. In the following sections, we will analyze the cutoff dependence of $Z$ and of the antipodal two-point function for the $\phi^2$ and $\phi^3$ flows in perturbation theory and compute the relevant counterterms. We stress that the problem of reducing cutoff effects in Hamiltonian truncation is important and well-explored, see e.g.~\cite[Sec.\@ VI]{truncRev} for a discussion of various methods. Recent work in this direction by various authors~\cite{Hogervorst:2014rta,Rychkov:2014eea,Rychkov:2015vap,Elias-Miro:2015bqk,Elias-Miro:2017tup,Elias-Miro:2017xxf} has culminated in extremely precise results for the $\phi^4$ flow on $\mbb{R} \times S^1$. Since there is no bona fide Hamiltonian on $S^d$, the approach we take is necessarily different from most of the literature, but will be similar in spirit to a perturbative method from Ref.~\cite{Giokas:2011ix}.

\subsection{Covariant results}

In the next sections, we will compute observables using time-dependent perturbation theory in order to derive their cutoff dependence. Working in the continuum limit $\La \to \infty$, the same observables may be computed using covariant methods. Here we briefly present the results of such a computation, performed using conformal perturbation theory in the case of the $m^2  = 0$ theory on $S^3$. We have
\bsub
\begin{align}
\ln \frac{Z(\la_2,\la_3)}{Z(0,0)} &=  \frac{\pi^2}{16}  \bar\la_2^2 + \frac{1}{192(3-d)} \bar\la_3^2 +  \ldots \label{eq:pff}\\
R^{d-2}\expec{\phi(\NN)\phi(\SS)}_\text{conn}&= \frac{1}{8\pi} - \frac{1}{4\pi}  \bar\la_2 + \frac{\pi^2-4}{16\pi} \bar\la_2^2 + \frac{2\mathtt{Catalan}-1}{16\pi^2} \bar\la_3^2 + \ldots \label{eq:NS0}\\
R^{d-2}\expec{\phi^2}_\text{conn} &= - \frac{1}{8} \bar\la_2 + \frac{1}{8} \bar\la_2^2 + \frac{1}{16\pi^2}  \bar\la_3^2 +  \ldots \label{eq:vevv}
\end{align}
\esub
omitting terms of order $\la_2^3$ and $\la_3^4$. The quantity $\mathtt{Catalan}$ denotes Catalan's constant, $\mtt{Catalan} \simeq 0.916$. The $1/(3-d)$ pole in the partition function reflects a logarithmic UV divergence that will be discussed later; all other terms in the perturbative expansion are finite. 

\subsection{Leading contribution to the partition function}
\label{sec:ct}

In what follows, we turn to the computation of the partition function for a general perturbation of the form~\reef{eq:int}.  The leading correction to $Z(\la)$ --- cf.\@ formula~\reef{eq:Zexp2} --- reads
\beq
\label{eq:int1}
\frac{Z(\la)}{Z(0)} = 1 + \bar{\la}^2  \int_{-\infty}^\infty\!\frac{d\tau_1}{(\cosh \tau_1)^{d}} \int_{-\infty}^{\tau_1}\!\frac{d\tau_2}{(\cosh \tau_2)^{d}}
\, \expec{\wh{\Oo}(\tau_1)\wh{\Oo}(\tau_2)}_0 + O(\bar\la^3)
\eeq
where $\wh{\Oo}(\tau)$ is the spatially integrated operator as defined in~\reef{eq:Ohat}. We are assuming that $\Oo$ is normal-ordered, such that $\expec{\Oo}_0$ vanishes. The two-point function $\expec{\wh{\Oo}(\tau_1)\wh{\Oo}(\tau_2)}_0$ does not have a universal form --- it's not constrained by any symmetries. However, at short distances we can approximate the full $\expec{\Oo \Oo}_0$ correlator by
\beq
\label{eq:OO2}
\expec{\Oo(\tau_1,\bn_1)\Oo(\tau_2,\bn_2)}_0 \; \limu{\msc{S} \to 0} \;  \frac{a}{(4R^2 \msc{S})^\DD} = a \left(\frac{\cosh \tau_1 \cosh \tau_2}{2R^2}\right)^\DD \frac{1}{(\cosh \tau_{12} - \bn_1\cdot\bn_2)^\DD} 
\eeq
for some constants $a>0$ and $\DD$,  writing $\tau_{12} \equiv \tau_1 - \tau_2$.\footnote{In passing, we remark that if the UV theory is a CFT and $\Oo$ has a well-defined scaling dimension $\DD$, then formula~\reef{eq:OO2} is exact. This can easily be checked by applying a Weyl transformation to~\reef{eq:OO2}. The choice $a = 1$ corresponds to a flat-space two-point function normalized as ${\expec{\Oo(x)\Oo(y)} = |x-y|^{-2\DD}}$.} In writing~\reef{eq:OO2} only the most singular term as $\msc{S} \to 0$ is taken into account, and any subleading terms $\propto 1/\msc{S}^{\DD'}$ with $\DD' < \DD$ are omitted. This approximation can be justified \emph{a posteriori}. In the theory of the massive boson with $\Oo = \phi^n$,  the coefficients $\DD$ and $a$ are given by
\beq
\DD = n(\th d-1)
\qaq
a = \frac{n!}{[(d-2)\mrm{S}_d]^n}
\eeq
provided that $d>2$. Notice that the coefficients $\DD$ and $a$ do not depend on the bare mass $m^2$, although all subleading terms do. In $d=2$ the approximation~\reef{eq:OO2} breaks down, since at short distances the correlator $\expec{\phi^n \phi^n}$ scales as $(\ln \msc{S})^n$.\footnote{It is perhaps possible to adapt the strategy of Refs.~\cite{Rychkov:2014eea,Rychkov:2015vap,Elias-Miro:2017tup} to deal with such correlators.} Integrating over angles in~\reef{eq:OO2}, we obtain
\beq
\label{eq:expl}
\expec{\wh{\Oo}(\tau_1)\wh{\Oo}(\tau_2)}_0 \stackrel{\tau_1 > \tau_2}{=} a\,   \Sd^2 \,  ( \cosh \tau_1 \cosh \tau_2 \, e^{-\tau_{12}} )^\DD \, {}_2F_1\!\left[{{\DD,\DD-d/2+1}~\atop~{d/2}} \Bigg| \, e^{-2 \tau_{12}} \right]\!.
\eeq
Working at finite cutoff means that only certain intermediate states are allowed, i.e.
\beq
\expec{\wh{\Oo}(\tau_1)\wh{\Oo}(\tau_2)}_0 \longrightarrow \sum_{\ket{\psi}} \braket{\emptyset}{\wh{\Oo}(\tau_1)}{\psi} \braket{\psi}{\wh{\Oo}(\tau_2)}{\emptyset},
\quad
E(\ket{\psi}) \leq \La.
\eeq
Eq.~\reef{eq:expl} does not refer to any explicit quantization. However, if we Taylor expand the hypergeometric function in~\reef{eq:expl}, the $k$-th term corresponds roughly to an intermediate state of energy $E \sim (\DD+2k)/R$ (taking into account the $\exp(-\DD \tau_{12})$ prefactor). Hence we can identify $\La \sim (\DD+2k_\text{max})/R$, at least for very large values of $\La$. This leads to the following estimate for $Z(\la)$ at cutoff $\La:$
\beq
\label{eq:summ0}
\frac{Z(\la)}{Z(0)} = 1 + a \bar\la^2   \mca{B}(\La) + \ldots,
\quad
\text{where}
\quad
\mca{B}(\La) =  \mrm{S}_d^2 \sum_{k=0}^{\La R/2} \frac{(\DD)_k (\DD - d/2+1)_k}{k!(d/2)_k} \, I_k
\eeq
and $I_k = I_k(\DD,d)$ is the following definite integral:
\beq
 \label{eq:Ikdef}
 I_k \ldef \int_{-\infty}^\infty\!\frac{d\tau_1}{(\cosh \tau_1)^{d-\DD}} \int_{-\infty}^{\tau_1}\!\frac{d\tau_2}{(\cosh \tau_2)^{d-\DD}} \, e^{-(\DD+2k)\tau_{12}}.
 \eeq
The integrals~\reef{eq:Ikdef} are convergent and can be evaluated in closed form, if desired. However, we're interested in the tail of the sum in~\reef{eq:summ0}, and therefore we only need the large-$k$ asymptotics of $I_k$:
\beq
\label{eq:Ikas}
I_k \stackrel{k \gg 1}{=} \frac{B(d-\DD,\th)}{d+2k} + O(1/k^2)
\eeq
where $B(x,y)$ is the Euler beta function --- this is a special case of Eq.~\reef{eq:G2as}.

We are now ready to determine the cutoff dependence of the sum~\reef{eq:summ0}. Using the asymptotics~\reef{eq:Ikas}, we find that the summand in~\reef{eq:summ0} scales as $k^{2\DD-d-1}$. This means that for $\DD < d/2$ the partition function is finite, whereas for $\DD \geq d/2$ it is divergent. The case $\DD > d/2$ will not be important in the present work. Treating the cases $\DD < d/2$ and $\DD = d/2$ separately, we obtain
\bsub
\begin{align}
&\DD < d/2: \label{eq:ccccc} \quad \mca{B}(\La) = 2^{d-2\DD-2} \mrm{S}_{d+1} \mrm{S}_d B(\th d-\DD,\th d) - \frac{\nu(\DD)}{(\La R)^{d-2\DD}} + O(1/\La^{d-2\DD+1}),\\
&\hspace{8cm}\nu(\DD) = \frac{ 2^{d-2\DD-2} \Sd^2 \Gamma(\th d)B(d-\DD,\th)}{(\th d-\DD)\Gamma(\DD)\Gamma(\DD-\th d+1)}\nn\\
&\DD = d/2: \quad \La \frac{d}{d\La} \mca{B}(\La) = \frac{(2\pi)^d}{\Gamma(d)} + O(1/\La).
\end{align}
\esub
In the finite case ($\DD<d/2$) we can compensate for the leading truncation error by shifting the Casimir energy as follows:
\bsub
\label{eq:ct1}
\beq
\DD < d/2: \quad S \, \mapsto \, S -  a \, \frac{\nu(\DD)}{\mrm{S}_{d+1}} \frac{\la^2}{ \La^{d-2\DD}} \int_{S^d}\sqrt{g}d^dx. \label{eq:rgi1}
\eeq
If $\DD=d/2$, a logarithmic counterterm {must} be added in order to make the partition function well-defined:
\beq
\label{eq:rgi2}
\hspace{7mm}\DD = d/2: \quad S \, \mapsto \, S + a (\Sd/2) \,  \la^2 \ln(\La/\mu)  \int_{S^d}\sqrt{g}d^dx
\eeq
\esub
where $\mu$ is an arbitrary energy scale. The only other dimensionful quantity (other than $R$) is $\la$, hence we can take $\mu = |\la|^{2/d}$ for definiteness. However, changing $\mu$ merely shifts the Casimir energy density by an amount proportional to $\la^2$.

Notice that both counterterms in~\reef{eq:ct1} are completely local. In general, it is not true that only local counterterms are generated in our scheme. For instance, if we attempt to further RG-improve the action, we will encounter terms that explicitly depend on the radius $R$. For instance, subleading terms in~\reef{eq:ccccc} are of the form $1/(\La R)^{d-2\DD+n}$, and only for even $n$ such a term can be compensated by adding a local counterterm to the action. Non-local counterterms are a generic feature of hard energy cutoffs, and they are discussed in more detail e.g.\@ in Refs.~\cite{Hogervorst:2014rta,Rychkov:2015vap,Elias-Miro:2017tup,Rutter:2018aog} in different settings.

Interactions with $\DD > d/2$ can be addressed in a similar fashion. For instance, in the 3$d$ $\phi^4$ theory, a linearly divergent counterterm must be added, similar to the logarithmically divergent counterterm from Eq.~\reef{eq:rgi2}.

\subsection{LO computation for  $\phi^2$ and $\phi^3$ interactions}

In the previous section, we provided a rough estimate of the cutoff dependence of the partition function, based on a general argument that did not refer to a specific action. Next, we will consider the $\phi^2$ and $\phi^3$ interactions on $S^d$ in the Hamiltonian picture of Sec.~\ref{sec:inter} and compute the cutoff dependence of $Z$ using time-dependent perturbation theory. The fact that both approaches agree provides a useful consistency check.

To leading order, the partition function $Z$ for a cubic theory with Lagrangian~\reef{eq:Sact} is given by
\beq
\ln \frac{Z(\la_2,\la_3)}{Z(0,0)} = \bar\la_2^2 \, \mca{C}_2(\La) + \bar\la_3^2 \, \mca{C}_3(\La) + O(\la_2^3,\la_3^4)
\eeq
The diagrams $\mca{C}_n(\La)$ are given by
\beq
\label{eq:cnexp}
\mca{C}_n(\La) = \frac{1}{n!} \sum_{\{\l_i\}} \Theta(\vareps(\l_1) + \ldots + \vareps(\l_n) \leq \La) A_n(\l_1,\ldots,\l_n) Q(\l_1,\ldots,\l_n)
\eeq
where
\beq
\label{eq:Andef}
A_n(\l_1,\ldots,\l_n) \ldef \int_{S^{d-1}}d\bn_1d\bn_2 \,\prod_{i=1}^n \frac{n_{\l_i}^d}{\mrm{S}_d} \; C_{\l_i}^d(\bn_1 \cdot \bn_2)
\eeq
and
\beq
Q(\l_1,\ldots,\l_n) \ldef \int_\mbb{R}\frac{d\tau}{(\cosh \tau)^d} \int_{-\infty}^\tau\!\frac{d\tau'}{(\cosh \tau')^d} \, \prod_{i=1}^n K_{\l_i}(\tau)K_{\l_i}(-\tau').
\eeq
The sum in~\reef{eq:cnexp} runs over all unordered tuples $\{\l_1,\ldots,\l_n\}$, but the $\Theta$ function only selects intermediate states with energy $E \leq \La$. The integrals $A_n$ can be performed analytically for all $n$ and $d$; closed-form expressions for $n=2,3$ are stated in Sec.~\ref{sec:harmonics}. On the other hand, the $Q(\l_i)$ integrals are rather complicated, at least for general values of $m^2$ and $d$, and as such must be computed numerically. 

Let us first consider $n=2$ in three dimensions. In this case we know that $\mca{C}_2(\La)$ has a finite limit as $\La \to \infty$. For definiteness, let's consider the massless case, for which the function $\mca{C}_2(\La)$ can be computed analytically. To wit
\begin{multline}
\label{eq:exp234}
\mca{C}_2(\La)\Big|_{m^2=0,d=3} = \sum_{\l=0}^\infty \Theta(2\vareps(\l) \leq \La) (\l + \th) Q(\l,\l),\\
Q(\l,\l)\Big|_{m^2=0,d=3} = \frac{2}{(2\l+1)^2} \left[{}_3F_2\!\left({{\;1,\;1,\;1}~\atop~{3,\l+\tfrac{3}{2}}}\,\Bigg|\,1 \right)-1 \right] \; \limu{\l \to \infty} \; \frac{1}{6\l^3}.
\end{multline}
In obtaining this result we make use of the integral $G_1(\a,\b)$ from Eq.~\reef{eq:G2sol}. Using the above asymptotics, it can be shown that 
\beq
\label{eq:c2exp}
\mca{C}_2(\La) \; \limu{\La \to \infty} \; \frac{\pi^2}{16} - \frac{1}{3\La R} + O(1/\La^2).
\eeq
According to Eq.~\reef{eq:rgi1}, the leading $1/\La$ truncation error in~\reef{eq:c2exp} can be compensated for by adding the following counterterm:
\beq
\label{eq:crj}
\dd S = - \frac{1}{6 \pi^2} \frac{\la_2^2}{\La} \int_{S^3} \sqrt{g}d^3x = - \frac{1}{3 \La R} \, \bar\la_2^2\, .
\eeq
Comparing Eqs.~\reef{eq:c2exp} and~\reef{eq:crj}, we find a perfect agreement between the general argument from the previous section and the specific computation in Eq.~\reef{eq:exp234}. 

We can treat the cubic ($n=3$) interaction in a similar fashion. Now, we have to consider the behavior of
\beq
\label{eq:C3la}
\mca{C}_3(\La) = \frac{1}{6} \sum_{\l_1 \l_2 \l_3} \Theta( \vareps(\l_1) + \vareps(\l_2) + \vareps(\l_3) \leq \La) A_3(\l_1,\l_2,\l_3) Q(\l_1,\l_2,\l_3)
\eeq
at large $\La$. The coefficient $A_3$ is computed in Eq.~\reef{eq:A3crj}, and in the massless limit  we furthermore have
\beq
\label{eq:Q3ex}
Q(\l_i)\Big|_{m^2=0,d=3} = \frac{2\pi}{(2\l_1+1)(2\l_2+1)(2\l_3+1)} \left[\pi \frac{(\tfrac{3}{2})_L^2}{L!^2} -(4L+3) \right]\!,
\quad
L = \th(\l_1 + \l_2 + \l_3).
\eeq
A precise determination of~\reef{eq:C3la} is somewhat difficult. However, it is easy to argue that $\mca{C}_3(\La)$ diverges logarithmically in $d=3$, as was already predicted in Eq.~\reef{eq:pff}. To prove this, consider the variation $\pd \mca{C}_3/\pd \La$ at large $\La$, such that the $\l_i$ can be treated as continuous variables. The sum over the $\l_i$ becomes an integral that localizes on a two-simplex $\l_1 + \l_2 + \l_3 \simeq \La R$ of volume $\sim (\La R)^2$. Using the explicit formula for $A_3$, we have $A_3(\l_i) \sim \La R$ on this simplex. Moreover, from~\reef{eq:Q3ex} it follows that $Q(\l_i) \sim 1/(\La R)^4$. Consequently,
\beq
d=3:\quad \frac{\pd \mca{C}_3(\La)}{\pd \La} \limu{\La \to \infty} (\La R)^2 \cdot (\La R) \cdot (\La R)^{-4}  = 1/(\La R)
\eeq
up to a multiplicative constant that we have not attempted to compute. Following the discussion from the previous section, this logarithmic divergence was to be expected as well, since in $d=3$ we have $\DD_{\phi^3} = 3/2 = d/2$. According to Eq.~\reef{eq:rgi2}, the theory can be made UV-finite by adding the following counterterm
\beq
\label{eq:e-m-t}
\dd S = \frac{1}{192 \pi^2} \ln(\La/|\la_3|^{2/3}) \la_3^2 \int_{S^3} \sqrt{g}d^3x = \frac{\bar\la_3^2}{96}  \ln(\La/|\la_3|^{2/3})
\eeq
which renormalizes $\mca{C}_3$. Consequently the subtracted quantity
\beq
\label{eq:C3pred}
\mca{C}_3^\text{ren}(\La) \ldef \mca{C}_3(\La) - \frac{1}{96} \ln(\La/|\la_3|^{2/3})
\eeq
should have a finite limit as $\La \to \infty$, as we have checked numerically up to large cutoffs ($\La R \sim 100$). 

Both in the case of the $\mca{C}_2$ and $\mca{C}_3$ diagrams in $d=3$, we estimated truncation errors up to those of order $1/\La$ resp.\@ $\ln \La$. In principle, it is possible to compute further RG-improvement counterterms by analyzing terms of order $1/\La^2$ resp.\@ $1/\La$.  Unlike the leading counterterms~\reef{eq:crj} and~\reef{eq:e-m-t}, all subleading counterterms will depend on the bare mass $m^2$. In order to compute such subleading counterterms in general, higher-order diagrams in perturbation theory must be taken into account as well. 

\subsection{Antipodal correlation function at (N)LO}
\label{sec:antip}

In the previous section, we considered the cutoff dependence of the partition function $Z$. Here we will turn our attention to the antipodal correlation function $\expec{\phi(\NN)\phi(\SS)}_\text{conn}$, using it to fix an additional counterterm. Working at finite cutoff $\La$, we have
\beq
\label{eq:fglc}
R^{d-2}\expec{\phi(\NN)\phi(\SS)}_\text{disc} = R^{d-2}G(\msc{S}=1)+ \bar\la_2 \, \mca{D}_{2,1}(\La) + \bar\la_2^2 \, \mca{D}_{2,2}(\La) + \bar\la_3^2 \, \mca{D}_{3,2}(\La) + \ldots
\eeq
omitting higher-order diagrams and disconnected diagrams. The free contribution
\beq
G(\msc{S}=1) = \frac{\ka^2}{\Sd R^{d-2}},
\quad
\ka \equiv \lim_{\tau \to \infty} K_0(\tau),
\eeq
does not depend on the cutoff scale $\La$.\footnote{This is a consequence of the North-South kinematics of the correlator in question: for a general two-point function $\expec{\phi(\tau_1)\phi(\tau_2}$, an infinite tower of intermediate states of the form $a_{\l j}^\dagger \ket{\emptyset}$ contributes to the Green's function~\reef{eq:prop2}. Yet in the limit where either $\tau_1 \to \infty$ or $\tau_2 \to -\infty$, only the state with $\l = j = 0$ contributes (since all other $K_\l$ vanish).} The same applies to the leading-order contribution $\mca{D}_{2,1}$, since
\beq
\label{eq:D21}
\mca{D}_{2,1}(\La) = - \frac{\kappa^2}{\mrm{S}_d} \int_\mbb{R} \frac{d\tau}{(\cosh \tau)^d} \, K_0(-\tau)K_0(\tau).
\eeq
Surprisingly, the next-to-leading term $\mca{D}_{2,2}$ is cutoff-insensitive as well. There are two different diagrams contributing to this quantity, yielding
\beq
\mca{D}_{2,2}(\La) = \frac{\kappa^2}{\Sd} \int_\mbb{R} \frac{d\tau}{(\cosh \tau)^d}\int^\tau_{-\infty} \frac{d\tau'}{(\cosh \tau')^d} \left[ K_0(-\tau)K_0(\tau)K_0(-\tau')K_0(\tau') + K_0(\tau)^2 K_0(-\tau')^2 \right].
\eeq
In all of the above cases, only intermediate states of energy $E = \vareps(0)$ or $E = 3\vareps(0)$ are propagated.

Finally, consider the leading contribution generated by the $\phi^3$ term. There are infinitely many intermediate states contributing, indexed by a spin $\l$, and only spins up to $\l_\text{max} \sim \La R/2$ are below the cutoff and contribute. Explicitly, we have
\begin{multline}
  \label{eq:fgl}
  \mca{D}_{3,2}(\La) = \half \frac{\kappa^2}{\Sd^2} \sum_{\l=0}^\infty n_{\l}^d \int_\mbb{R} \frac{d\tau}{(\cosh \tau)^d}\int^\tau_{-\infty} \frac{d\tau'}{(\cosh \tau')^d}  \,  K_\l(\tau)^2 K_\l(-\tau')^2 \\
  \times \left[\Theta( 2\vareps(\l) \leq \La) K_0(-\tau) K_0(\tau') + \Theta(2\vareps(\l) + 2\vareps(0) \leq \La) K_0(\tau)K_0(-\tau')\right].
\end{multline}
The two terms in~\reef{eq:fgl} correspond to diagrams with different time-orderings. In the massless limit the sum~\reef{eq:fgl} can be computed in closed form using the integral~\reef{eq:G2sol}. Setting $d \to 3$ this yields
\begin{multline}
\label{eq:D32sum}
\mca{D}_{3,2}(\La)\Big|_{m^2 = 0,d=3} = \frac{1}{32\pi} \sum_{\l} \frac{\Theta( 2\vareps(\l) \leq \La)}{2\l+1} \!\left[\frac{4}{\pi}\frac{\l!^2}{(\th)_\l^2} -4\l-1 \right] \\+\frac{\Theta( 2\vareps(\l) + 2\vareps(0) \leq \La)}{2\l+1}\!\left[ \pi \frac{ (\tfrac{3}{2})_\l^2}{ \l!^2} -4\l-3 \right].
\end{multline}
Evaluating~\reef{eq:D32sum} in the limit of large $\La$, we find
\beq
\label{eq:NSerr}
(\ldots)  = \frac{2\mtt{Catalan}-1}{16\pi^2} - \frac{1}{128 \pi \La R} + O(1/\La^2)
\eeq
consistent with~\reef{eq:NS0}. We would like to compensate for this $1/\La$ truncation error by adding a (local) counterterm proportional to $\phi^2$:
\beq
\label{eq:phi2ct}
\dd S = - c_3\,  \frac{\la_3^2}{\La} \int_{S^3}\sqrt{g}d^dx\, \phi^2(x)
\eeq
where $c_3$ is a dimensionless constant. The counterterm~\reef{eq:phi2ct} shifts the $\expec{\phi(\NN)\phi(\SS)}$ two-point function by an amount
\beq
\dd \expec{\phi(\NN) \phi(\SS)} =   \frac{c_3}{2\pi} \frac{ \bar\la_3^2}{\La R^2} + O(1/\La^2).
\eeq
By comparing with~\reef{eq:NSerr}, we conclude that setting $c_3=1/64$ eliminates the leading cutoff error completely.

Ultimately, we are interested in computing the \emph{connected} antipodal two-point function. So far we have been cavalier about disconnected contributions to the $\expec{\phi(\NN)\phi(\SS)}$ correlator in Eq.~\reef{eq:fglc}. At order $\la_3^2$, we have discarded one disconnected diagram. Had we used a fully local regulator, then this disconnected diagram would cancel after dividing by the partition function $Z(\la_3)/Z(0)$. Unfortunately, in our hard-cutoff scheme the contribution of the two diagrams does not cancel entirely. There is a spurious contribution $\wt{\mca{D}}$ to the antipodal correlation function at order $\la_3^2$:
\begin{multline}
\label{eq:Dspur}
R^{d-2}\expec{\phi(\NN)\phi(\SS)}_\text{conn} = (\text{connected diagrams}) + \bar\la_3^2 \, \wt{\mca{D}}(\La) + O(\la_3^4),\\ 
\wt{\mca{D}}(\La) = -\frac{\ka^2}{6\Sd}  \sum_{\{\l_i\}} \Theta(\La - 2\vareps(0) \leq \vareps(\l_1)+\vareps(\l_2)+\vareps(\l_3) < \La) A_3(\l_1,\l_2,\l_3) Q(\l_1,\l_2,\l_3).
\end{multline}
In order to reproduce the correct continuum limit~\reef{eq:NS0}, it is crucial that $\wt{\mca{D}}(\La)$ vanishes as $\La \to \infty$. This is not immediately obvious. By comparing the summand of~\reef{eq:Dspur} to Eq.~\reef{eq:cnexp}, it is however easy to see that
\beq
\wt{\mca{D}}(\La) = -\vareps(0)\frac{2\ka^2}{\Sd} \, \frac{\pd \mca{C}_3(\La)}{\pd \La} 
\eeq
up to errors that can be neglected when $\La \gg \vareps(0)$. Since in $d=3$ the function $\mca{C}_3$ grows logarithmically with $\La$, we conclude that $\wt{\mca{D}}(\La)$ vanishes as $1/\La$ as the cutoff is removed, as desired.

\section{Nonperturbative results}\label{sec:numerics}

In the first half of this paper, a method to compute observables on $S^d$ was proposed for scalar QFTs on the $d$-sphere, and subsequently the leading counterterms generated by $\phi^2$ and $\phi^3$ interactions were computed. In what follows, we will test our method numerically, i.e.\@ in the strong coupling regime. First, we explain in more detail how the computation in question is performed. Second, we turn to the $\phi^2$ interaction on $S^3$, where our numerical data can be compared to analytic formulas. Finally, we consider the $i \phi^3$ interaction, where we will check that the counterterm prescription from Sec.~\ref{sec:cutoff} renders the theory UV-finite. 

\subsection{Implementation}
\label{sec:implementation}

The framework used to perform nonperturbative computations in our scheme has been explained in Secs.~\ref{sec:hamont} and~\ref{sec:truncgen}. Here, we will provide additional details that are needed to reproduce our results. Some non-essential comments are discussed in Appendix~\ref{sec:remarks}. Suppose that the spacetime dimension $d$, the bare mass $m^2$  and the cutoff $\La$ are fixed. Then any computation proceeds in three steps, schematically:
\begin{enumerate}
\item Generate a basis of all Fock space states $\ket{\psi_i}$ with energy $E(\ket{\psi_i})$ below the cutoff $\La$;
\item Generate  matrices $V_n(\tau) \equiv [V_n(\tau)]\ud{j}{i}$ defined as
  \[
   R^{n(d/2-1)} \int_{S^{d-1}}\!d\bn\, \phi^n(\tau,\bn)\ket{\psi_i} \rdef \sum_j [V_n(\tau)]\ud{j}{i} \ket{\psi_j} + \text{states above the cutoff};
  \]
\item Compute observables using the Trotter formula~\reef{eq:trotter}. 
\end{enumerate}

Let us provide some additional details regarding these three steps. Concerning the first step, the reader will remark that the total number of Fock space states grows rapidly (exponentially) with the cutoff $\La$. Listing \emph{all} Fock space states below the cutoff $\La$, it would be prohibitive to store the matrices $V_n(\tau)$ in memory if $\La R \gtrsim 15$. However, the Dyson operator $U$ is invariant under $\mrm{O}(d)$ transformations acting on the spherical coordinate $\bn$, i.e.\@ both under $\SO(d)$ transformations and a parity transformation (if $d$ is odd, parity acts as $\bn \to -\bn$). In the case of the partition function~\reef{eq:Udef1} and the antipodal two-point function~\reef{eq:NS1}, all in- and out-states are $\mrm{O}(d)$ scalar states. Consequently, only $\mrm{O}(d)$ scalar intermediate states are needed, which strongly constrains the list of states $\ket{\psi_i}$ that must be taken into account.

Let's specialize to $d=3$, denoting the generators of $\SO(3)$ as $L_\pm$ and $L_z$. It is easy to generate a basis of parity-even states obeying $L_z \ket{\psi} = 0$, and it tedious but straightforward to select all states that in addition obey $L_\pm \ket{\psi} = 0$. For various values of $\La$, a counting of the dimension of the Fock space $\msc{F}(\La)$ for $d=3$ and $m^2 = 0$ is shown in Table~\ref{table:count}. (If $\la_3 = 0$, the action is invariant under a $\mbb{Z}_2$ global symmetry $\phi \to -\phi$ which further constrains the Fock space.)
\begin{table}[htbp]
  \begin{center}
\begin{tabular}{c | r | r | r}
  $\La R$ & all states & $L_z = 0$ \& parity-even & scalars \\
  \hline
  10 & 6057 & 422 & 58 \\
  15 & 193155 & 9231 & 439 \\
  20 & 4425606 & 166802 & 3782
\end{tabular}
\end{center}
\caption{Counting the number of Fock space states obeying $\mrm{O}(d)$ selection rules in $d=3$ for $m^2=0$, for various values of the cutoff $\La$.}
\label{table:count}
\end{table}

The second step, i.e.\@ the computation of the matrices $V_n(\tau)$, is again tedious but straightforward. The normal-ordered operators $\phi^2$ and $\phi^3$ are defined explicitly in Eq.~\reef{eq:noop}. Consequently, the action of either of these operators on a Fock space state $\ket{\psi_i}$ is completely determined by the canonical commutation relations. Note that the matrix elements $[V_n(\tau)]\ud{j}{i}$ depend on $\tau$ only through the functions $K_\l(\tau)$ and $K_\l(-\tau)$. 

The computation of observables using the Trotter formula is straightforward. Consider for instance the partition function~\reef{eq:Udef1}, given couplings $\la_n$ and a number of timeslices $T$. We can directly use formula~\reef{eq:trotter} to compute an estimate for $Z(\la_n)/Z(0)$. In pseudocode, our algorithm reads:
 \begin{align*}
   &\ket{\psi} \longleftarrow \ket{\emptyset}\\
   &\mathtt{for} \; k=0,\ldots,T-1:\\
   &\qquad z \longleftarrow (k+\th)/T \\
   &\qquad \ket{\psi} \longleftarrow  \ket{\psi} - \frac{\mrm{S}_{d+1}}{\mrm{S}_d T} \,  \sum_n \frac{\bar\la_n}{n!} \,  V_n(\tau(z)) \ket{\psi}\\
   &\mathtt{out} \longleftarrow \brakket{\emptyset}{\psi}.
 \end{align*}
If counterterms are added to the action, the fourth line needs to be modified in an obvious way.  With $\ket{\psi}$ we denote an $N$-dimensional vector, where $N$ is the number of scalar Fock space states. Above we have used the $z$-coordinate introduced in Eq.~\reef{eq:zdef}, as well as the Jacobian
 \[
 \frac{\dd \tau}{(\cosh \tau)^d} = \frac{1}{z'(\tau)} \frac{\dd z}{(\cosh \tau(z))^d} = \frac{\mrm{S}_{d+1}}{\mrm{S}_d} \, \dd z.
 \]
 Since times are uniformly sampled in $z$, it follows that $\dd z = 1/T$.
 
 We have performed all of these steps in $\mathtt{Mathematica}$ on a laptop computer. The only subtlety in implementing the above algorithm has to do with numerical precision. The algorithm in question performs a huge number of floating point operations (addition and multiplication). Given a truncated Fock space of dimension $N$ and $T$ timeslices, an estimate of the number of floating point operations is $N_\text{op} = N^2 T$, which for $\La R = 20$ and $T = 2500$ evaluates to $N_\text{op} \sim 4 \cdot 10^{10}$. Consequently, using $\mtt{MachinePrecision}$ does not always yield satisfactory results, and typically it is necessary to work with arbitrary-precision numbers. The number of digits required depends on $T$, $\La$ and the couplings $\bar\la_n$; in this work we have used up to 300 digits (out of an abundance of caution). 

 Finally, let us make a remark about the discretization of the Trotter formula, i.e.\@ the number $T$ of timeslices that is used. In practice, we repeat the same computation of an observable $f$ for a range of values $T$ up to some $T_\text{max}$, labeling the data points as $f(T)$. Next, we verify that the $1/T$ falloff of the discretization error is satisfied, which allows for an extrapolation to $T = \infty$. This procedure yields an estimate of the observable $f(T)$ as $T \to \infty$ together with an error estimate $\dd f_\text{disc} = |f(\infty) - f(T_\text{max})|$. 
 
\subsection{$\phi^2$ flow on $S^3$}

Let us first consider the case where we only turn on a $\phi^2$ interaction on $S^3$ with coupling $\la_2$. This is nothing but the Gaussian action of a boson with a physical mass $M^2 = m^2 + \la_2$. The partition function on $S^d$ can be computed exactly, yielding:\footnote{A simple way to obtain~\reef{eq:Fc} is through the identity
  \[
  \frac{d}{d\la} \ln \frac{Z(\la)}{Z(0)} = - \mrm{S}_{d+1} R^d \expec{\Oo}_{\text{conn},\la}
  \]
  which holds for a general perturbation of the form~\reef{eq:int}. In the case at hand, $\expec{\Oo} = \expec{\phi^2}$ can be extracted from the short-distance behaviour of the Green's function~\reef{eq:prop1}.
  }
\beq
\label{eq:Fc}
\ln \frac{Z(M)}{Z(m)} = \half \int_{m^2R^2}^{M^2R^2}\!dx\! \left[q(x)-q(m^2R^2)\right]
\quad
\text{where}
\quad
q(x) = \frac{\pi}{2}\frac{\sqrt{\tfrac{1}{4}-x}}{ \tan \pi \sqrt{\tfrac{1}{4}-x}}.
\eeq
Likewise, the antipodal two-point function $\expec{\phi(\NN)\phi(\SS)}_\text{conn}$ correlator can be obtained from the Green's function~\reef{eq:prop1}, namely
\beq
\label{eq:antiex}
\expec{\phi(\NN)\phi(\SS)}_\text{conn} = \frac{1}{4\pi^2 R} \, \Gamma\big(1+\sqrt{\tfrac{1}{4}-M^2R^2}\big)\Gamma\big(1-\sqrt{\tfrac{1}{4}-M^2R^2}\big) \;
\limu{R \to \infty} \; \frac{M}{2\pi} \exp(-\pi M R).
\eeq
The large-$R$ behavior of~\reef{eq:antiex} simply shows that the correlation length of the massive boson equals $\xi = 1/M$, since the geodesic distance between the two poles equals $\pi R$.

\begin{figure}[htb]
\begin{center}
  \includegraphics[scale=1]{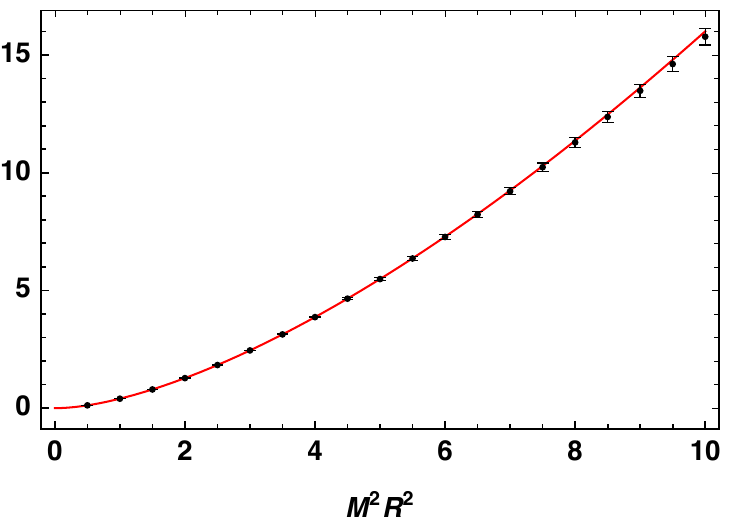}
  \hspace{5mm}
  \includegraphics[scale=1]{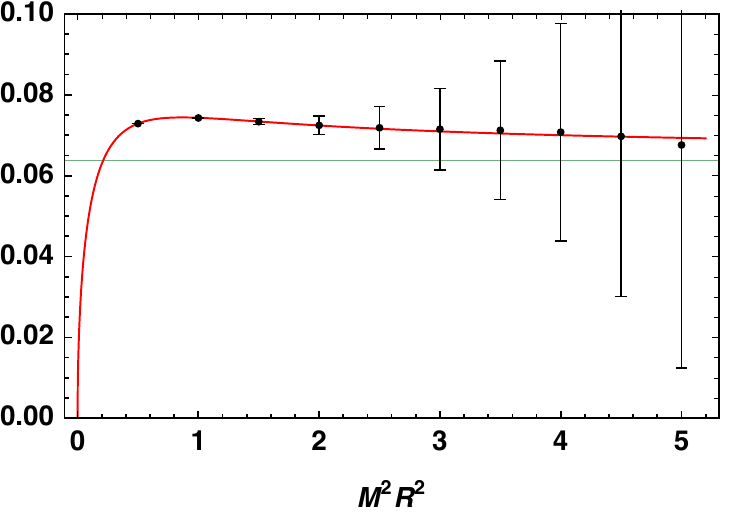}
\end{center}
\caption{{\bf Left}: the log of the partition function, $\ln Z(M)/Z(0)$, for the $\th M^2 \phi^2$ flow~\reef{eq:actpr}, scanning over a range of values of $M^2R^2$ (horizontal axis). Black dots with error bars: numerical data; {\Red red} curve: exact result. {\bf Right}: the same observable, recomputed using the renormalized action $S_\text{ren}$ from Eq.~\reef{eq:newact}. {\Green Green} horizontal line: $F_\text{scalar}$, the $M^2R^2 \to \infty$ limit of the free energy.}
\label{fig:exppart}
\end{figure}
We will compare the exact predictions~\reef{eq:Fc} and~\reef{eq:antiex} to numerical data obtained using our scheme, setting $m^2 = 0$ from now on. After adding the RG improvement term~\reef{eq:crj}, the action reads
\beq
\label{eq:actpr}
S = S_0+ \int_{S^3}\!\sqrt{g}d^3x\!\left[ \half M^2 \phi^2(x) - \frac{M^4}{6\pi^2 \La}  \right]\!.
\eeq
We will use values of $T \leq T_\text{max} = 2500$ and $\La R \leq 20$, scanning over couplings up to $M^2 R^2 = 10$. For a given coupling $M^2 R^2$, we expect and observe a truncation error that decreases as $1/\La^2$. Consequently, we can extrapolate the numerical data to $\La = \infty$, assigning error bars based on this extrapolation. In Fig.~\ref{fig:exppart} (left), we plot our numerical results and compare them to Eq.~\reef{eq:Fc}. For all radii $R$ in this range we find agreement within error bars, with relative errors at the $\%$ level or better. This is an excellent consistency check of our method, because the theory is strongly coupled in this regime. To make this precise, the observable $\ln Z(M)$ admits an asymptotic perturbative expansion $\sum_{n} \a_n (M^2R^2)^n$ with coefficients $\a_n$ that are sign-alternating and that grow exponentially with $n$. Therefore perturbation theory is a poor approximation to the exact result unless $M^2R^2 \ll 1$.

The $\phi^2$ theory flows from the free boson CFT in the UV to a gapped (empty) QFT in the IR. After adding two curvature counterterms from Eq.~\reef{eq:3dZ}, we expect that the free energy in the limit $MR \to \infty$ asymptotes to the $F$-coefficient of the free scalar CFT, since the empty QFT has $F = 0$. The necessary counterterms can be extracted from formula~\reef{eq:Fc} --- see for instance the discussion in Appendix A.1 of~\cite{Klebanov:2011gs}. Consequently, if we recompute $Z(M)$ using the renormalized action\footnote{The same flow was recently discussed in~\cite{Ghosh:2018qtg} in the context of putative $F$-functions that monotonically interpolate between UV and IR fixed points.}
\beq
\label{eq:newact}
S_\text{ren} = S +  \int_{S^3}\sqrt{g}d^3x\! \left[  \frac{M^3}{12\pi} - \frac{M}{192\pi} \Ricci \right]
\eeq
we expect that
\beq
\label{eq:Fasss}
\ln \frac{Z(M)}{Z(0)} \, \limu{R \to \infty} \, F_\text{scalar}  +  O(1/MR),
\quad
F_\text{scalar} = \frac{\ln 2}{8} - \frac{3\zeta(3)}{(4\pi)^2} \, \simeq \, 0.0638.
\eeq
The value of $F_\text{scalar}$ is usually derived using zeta function regularization, see e.g.~\cite{Klebanov:2011gs}.  In the right plot of Fig.~\ref{fig:exppart} we compare the numerical data to the prediction~\reef{eq:Fasss}. For $M^2R^2 \lesssim 5$ the data are in good agreement with $F_\text{scalar}$. For larger values of $M^2R^2$ the error estimates grow rapidly, and there is no longer a meaningful comparison between the numerical data and $F_\text{scalar}$. The error estimates in the above plot, based on the extrapolation from $\La R = 20$ to $\La = \infty$, are clearly rather conservative.

\begin{figure}[htbp]
\begin{center}
  \includegraphics[scale=1]{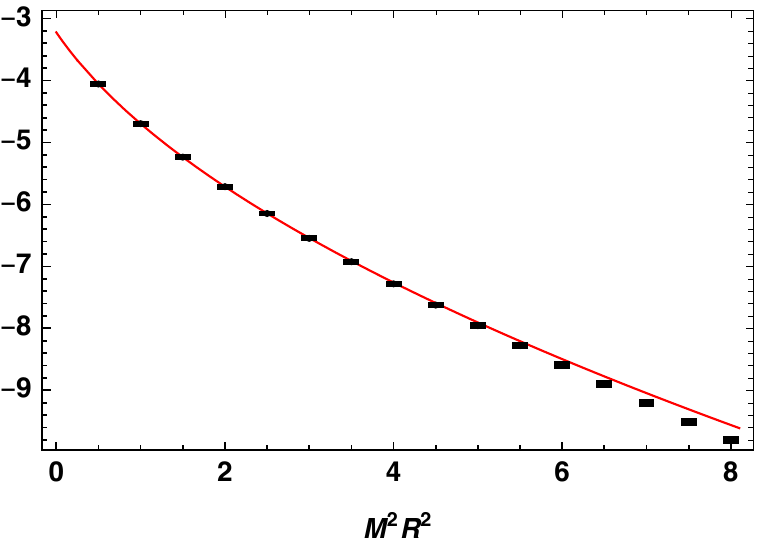}
\end{center}
\caption{{The logarithm of the correlator $\expec{\phi(\NN)\phi(\SS)}_\text{conn}$ in the presence of the $\th M^2 \phi^2$ flow, scanning over a range of couplings $M^2 R^2$ (horizontal axis) at cutoff $\La R = 20$. Black dots with error bars: numerical data; {\Red red} curve: exact result from Eq.~\reef{eq:antiex}. }}
\label{fig:expvev}
\end{figure}

Finally, we can compute the antipodal two-point function $\expec{\phi(\NN)\phi(\SS)}$ in the case of the $\phi^2$ flow. In Fig.~\ref{fig:expvev} we compare formula~\reef{eq:antiex} to numerical data computed in our scheme. Since the correlator in question decays exponentially with $MR$, it is more convenient to plot the observable
\[
f(MR) \ldef \ln R^{d-2} \expec{\phi(\NN)\phi(\SS)}_\text{conn}
\]
which decays linearly with $M R$. The plot in question is measured at cutoff $\La R = 20$, and we have not extrapolated to $\La = \infty$, contrary to the plots from Fig.~\ref{fig:exppart}; the error bars only reflect the extrapolation to $T = \infty$, cf.\@ the discussion at the end of Sec.~\ref{sec:implementation}. Up to $M^2 R^2 \simeq 5$, we find excellent agreement between our data and the exact formula; for larger radii, it is necessary to use higher cutoffs. 

\subsection{Cubic interaction}

\begin{figure}[htb]
\begin{center}
  \includegraphics[scale=1]{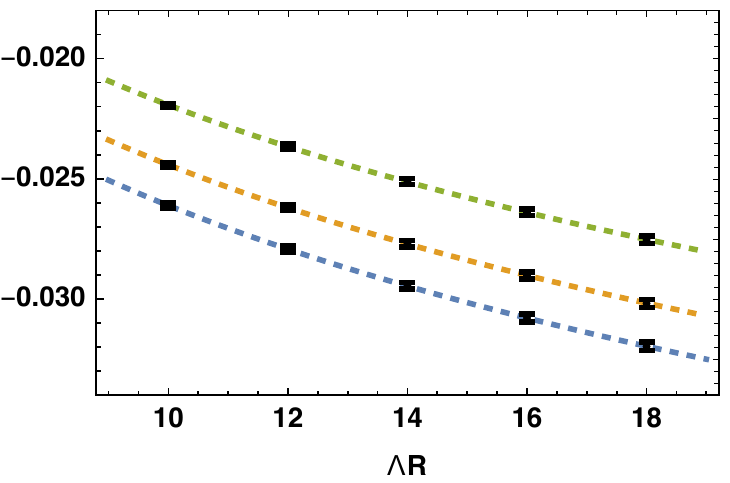}
  \hspace{5mm}
  \includegraphics[scale=1]{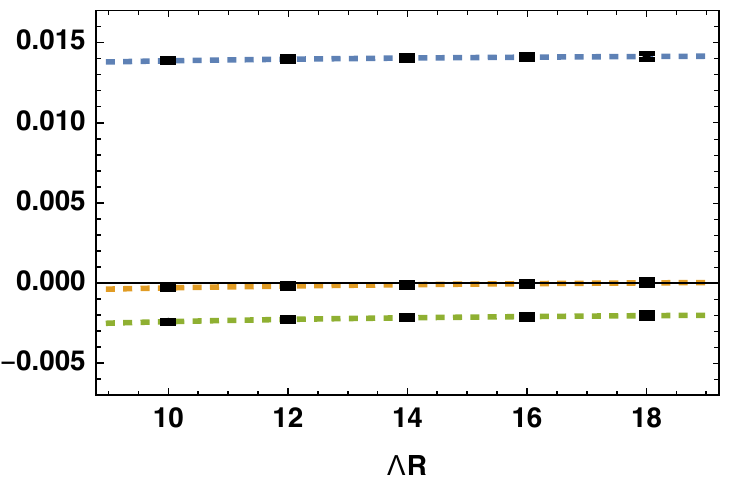}
\end{center}
\vspace{-5mm}
\caption{{\bf Left}: Logarithm of the partition function $h(\bar\la)$ for the bare action $S_\text{bare}$ as a function of the cutoff $\La R$ (horizontal axis). Black dots with error bars: numerical data; dotted lines: logarithmic fits. The three curves correspond to different couplings $\bar\la = \{0.1,1,2\}$, shown in \{{\Blue blue}, {\Orange orange}, {\Green green}\}.  {\bf Right:} same observable, computed for the renormalized action $S_\text{ren}$. The dotted lines now correspond to $1/\La$ fits.}
\label{fig:cubpart}
\end{figure}

Next, we consider a cubic interaction on $S^3$ with imaginary coupling
\beq
\label{eq:bare}
S_\text{bare} = S_0 +  \frac{i \la}{3!} \int_{S^3} \sqrt{g}d^3x  \, \phi^3(x)
\eeq
again using the $m^2 = 0$ theory as a starting point. We will consider the following values of the coupling: $\bar\la \equiv \la R^{3/2} = \{0.1,1,2\}$, so we can test both the perturbative and the strong-coupling regime. As discussed at length in Sec.~\ref{sec:cutoff}, the bare action is UV-divergent, whereas the renormalized action
\beq
S_\text{ren} = S_\text{bare} + \int_{S^3} \sqrt{g}d^3x \left[ - \frac{\la^2}{192\pi^2} \ln(\La/|\la|^{2/3}) + \frac{\la^2}{64 \La} \phi^2(x) \right]
\eeq
is expected to have a well-defined continuum limit. We will test this prediction by scanning over a range of cutoffs, from $\La R = 8$ to $\La R = 18$. To be precise, we show plots of the quantity
\[
h(\bar\la) \ldef \bar\la^{-2} \ln \frac{Z(\la)}{Z(0)}
\]
as a function of the cutoff $\La$, computed for the two actions $S_\text{bare}$ and $S_\text{ren}$, on the left (resp.\@ right) side of Fig.~\ref{fig:cubpart}. Indeed we observe that the bare partition function is divergent, whereas the free energy of the renormalized theory has a finite limit as $\La \to \infty$ up to an error of $1/\La$. 

\begin{figure}[htb]
\begin{center}
  \includegraphics[scale=1]{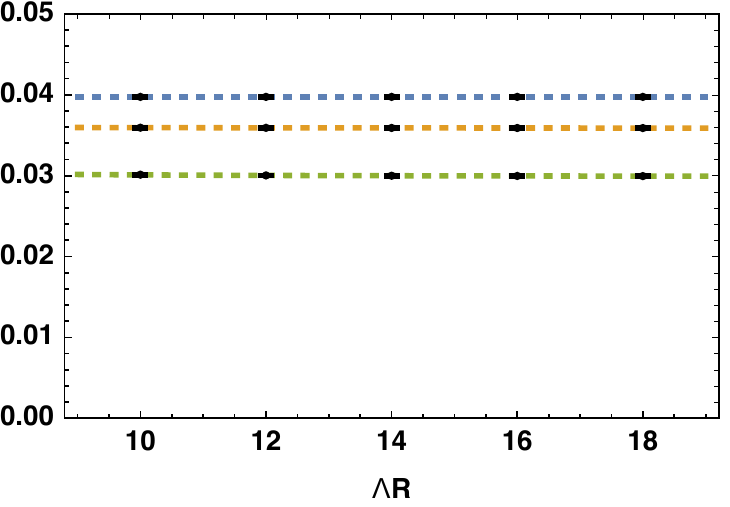}
\end{center}
\vspace{-5mm}
\caption{The antipodal correlator $R\, \expec{\phi(\NN)\phi(\SS)}$ for the cubic action $S_\text{ren}$ as a function of the cutoff $\La$ (horizontal axis). The three different curves correspond to the different couplings $\bar\la$, see Fig.~\ref{fig:cubpart}. Black points with error bars: numerical data. Dotted lines: $1/\La^2$ fits. }
\label{fig:cubap}
\end{figure}

As in the case of the $\phi^2$ flow, we also compute the antipodal correlation function $\expec{\phi(\NN)\phi(\SS)}_\text{conn}$. The results are shown in Fig.~\ref{fig:cubap}; we find that the measured correlator converges rapidly, roughly as $1/\La^2$, to its value in the continuum limit. This observation is agreement with the discussion from Sec.~\ref{sec:antip}.\footnote{In Sec.~\ref{sec:antip} the presence of a spurious term vanishing as $1/\La$ was mentioned. This term appears with a small coefficient, of order $10^{-2} \cdot \bar\la^2$ relative to the tree level contribution. For sufficiently large $\La$, we therefore expect a $1/\La$ decay of the truncation error. To properly analyze this issue, more data are needed.}

\section{Ward identities and one-point functions}
\label{sec:onept}

The cutoff $\La$ used in our scheme breaks the full rotation symmetry $\SO(d+1)$ of the $d$-sphere, although a subgroup $\SO(d)$ of spatial rotations is preserved. This translates to violations of the Ward identities~\reef{eq:Wardid} at the cutoff scale. It is crucial that the full spacetime symmetry group is recovered in the limit $\La \to \infty$. In many cases (e.g.\@ lattice regularizations) the regulator used is \emph{local}, and general RG arguments can be used to argue that spacetime symmetries are restored in the continuum limit. Unfortunately the regulator $\La$ is nonlocal, hence there is no simple classification of $\SO(d+1)$-violating counterterms. However, on a case-by-case basis we can measure violations of Ward identities to better understand the restoration of rotation invariance. In this section, we specialize to perhaps the simplest example, the one-point function of $\expec{\phi^2(\tau,\bn)}$ in the presence of quadratic and cubic interactions on $S^3$, computing violations of the Ward identity to second order in perturbation theory.

\subsection{Operator renormalization at leading order}
\label{sec:leador}

As in Sec.~\ref{sec:ct}, let's start by considering a general perturbation $\Oo$ in $d$ dimensions. The one-point function $\expec{\Oo(\tau,\bn)}$ is generated at first order in the coupling $\la$, to wit
\beq
\expec{\Oo(\tau,\bn)} = -\bar\la \int_{-\infty}^{\infty}\!\frac{d\tau'}{(\cosh \tau')^d} \, \braket{\emptyset}{\mrm{T}\Oo(\tau,\bn)\wh\Oo(\tau')}{\emptyset}_0 + O(\la^2).
\eeq
As discussed in Sec.~\ref{sec:ct}, the $\expec{\Oo \Oo}_0$ correlator is non-universal, and to simplify the following computation we use the short-distance approximation introduced in Eq.~\reef{eq:OO2}. Using the same logic as in that section, we arrive at the following estimate for the VEV $\expec{\Oo}$:
\begin{multline}
\label{eq:oneptsum}
\expec{\Oo(\tau,\bn)}  = -   \frac{\bar\la a}{R^\DD} \, \mca{V}(\tau,\La) + \ldots,\\
\quad
\mca{V}(\tau,\La) = \Sd \sum_{k=0}^{\La R/2}  \frac{(\DD)_k (\DD - d/2+1)_k}{k!(d/2)_k}  \; (\cosh \tau)^\DD \, \wh{G}_0(d-\DD,\DD+2k|\tau).
\end{multline}
Here $\wh{G}_0$ is a special function, defined in~\reef{eq:G1alt}. The upper limit $k_\text{max} = \La R/2$ in~\reef{eq:oneptsum} is a rough estimate, just as in Sec.~\ref{sec:ct}, but it suffices since we are only interested in the leading scaling with $\La$.
For $\DD < d/2$, the expansion~\reef{eq:oneptsum} converges uniformly in $\tau$, whereas for $\DD \geq d/2$ it diverges, reflecting a UV divergence of the VEV $\expec{\Oo}$.  Henceforth we will assume that $\DD < d/2$.

We claim that the Ward identity for $\expec{\Oo}$ (which must be constant) is violated by effects of order $1/\La^{d-2\DD}$. To make this precise, we have to analyze the behaviour of $\mca{V}(\tau,\La)$ at large $\La$. This can be done using Eq.~\reef{eq:G1as}, which leads to the estimate
\begin{multline}
\label{eq:V0as}
\mca{V}(\tau,\La) =   2^{d-2\DD-1} \mrm{S}_d B(\th d-\DD,\th d) -  \frac{\rho(\DD)}{(\La R \cosh \tau)^{d-2\DD}} + O(1/\La^{d-2\DD+1}),\\
\rho(\DD) = \frac{2^{d-2\DD} \Gamma(\th d)}{(d-2\DD)\Gamma(\DD)\Gamma(\DD-\th d+1)}.
\end{multline}
Indeed, the leading truncation error depends explicitly on $\tau$. The above error term can be canceled by renormalizing the operator $\Oo$ by a nonlocal counterterm. Indeed, suppose that we define a renormalized operator $\Oo_\text{r}(\tau,\bn)$ as follows:
\beq
\label{eq:Oren}
 \Oo(\tau,\bn) \rdef \Oo_\text{r}(\tau,\bn) + a\la \, \frac{ \rho(\DD)}{(\La \cosh \tau)^{d-2\DD}} \unit .
 \eeq
Then it is easy to see $\expec{\Oo_\text{r}(\tau,\bn)}$ will be less cutoff-sensitive, in the sense the error term of order $1/\La^{d-2\DD}$ in~\reef{eq:V0as} will be absent. In principle it's possible to compute subleading truncation effects as well, but we have not done so in the present work. 

 \subsection{(N)LO computation for $\expec{\phi^2}$}
 
\begin{figure}[htbp]
\begin{center}
  \includegraphics[scale=1]{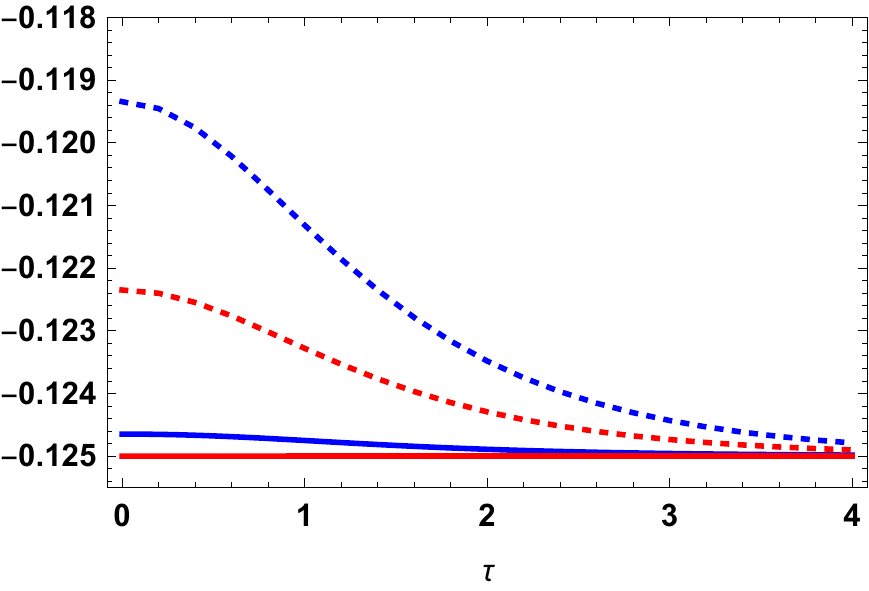}
\end{center}
 \caption{The leading-order contribution $\mca{E}_{2,1}$ to the VEV $\expec{\phi^2(\tau)}$ in the massless 3$d$ theory, as a function of $\tau$. We show the result for two different cutoffs: $\La R = 15$ ({\Blue blue}) and $\La R = 30$ ({\Red red}). The dotted (resp.\@ solid) line corresponds to VEV of the bare (renormalized) operator. The exact value of this diagram is $\mca{E}_{2,1}(\La \to \infty) = -1/8$, see Eq.~\reef{eq:vevv}.}
\label{fig:phi2vev}
\end{figure}

 We can test this prescription in the case of the $\phi^2 + \phi^3$ theory on $S^3$, taking $\Oo = \phi^2$. To proceed, we compute the VEV $\expec{\phi^2(\tau,\bn)}$ for the bare operator $\phi^2$ using time-dependent perturbation theory, which yields
 \beq
 \label{eq:vevmic}
  R^{d-2} \expec{\phi^2(\tau,\bn)}_\La =  \bar\la_2 \, \mca{E}_{2,1}(\tau,\La) + \bar\la_2^2 \, \mca{E}_{2,2}(\tau,\La)+ \bar\la_3^2 \, \mca{E}_{3,2}(\tau,\La) + O(\la_2^3,\la_3^4)
  \eeq
  for some functions $\mca{E}_{n,k}(\tau,\La)$ that will be displayed later. 
  Let us focus on the leading-order contribution:
  \beq
  \label{eq:E21}
  \mca{E}_{2,1}(\tau,\La) =   - \frac{1}{\Sd} \sum_{\l} n_{\l}^d\, \Theta(2\vareps(\l) \leq \La)  \left[\int_{-\infty}^\tau\!\frac{d\tau'}{(\cosh \tau')^d}  \left[K_\l(\tau)K_{\l}(-\tau')\right]^2  + (\tau \to -\tau)\right].
  \eeq
  According to the discussion from the previous section, we expect that $\mca{E}_{2,1}(\tau,\La)$ depends on $\tau$ through effects of order $1/\La^{d-2\DD}= 1/\La$. Moreover, Eq.~\reef{eq:Oren} provides a quantitative prediction, namely that the VEV of the renormalized operator
   \beq
 \label{eq:phict}
 d=3: \quad \phi^2_\text{r} \ldef \phi^2 -  \frac{\la_2}{4\pi \La \cosh \tau} \unit
 \eeq
 violates the Ward identity only through subleading effects of order $1/\La^2$, at least to leading order in $\la_2$. We have tested these predictions for the $m^2 = 0$ theory by evaluating the function~\reef{eq:E21} at large values of $\La$.  In Fig.~\ref{fig:phi2vev} we compare the VEV of the bare operator $\phi^2$ to the one of the renormalized operator $\phi^2_\text{r}$ for two values of the cutoff $\La$ at first order in $\la_2$. Although rotation-invariance breaking effects vanish as $1/\La$, they are clearly observable for small values of $\La$ in the case of the bare operator. As expected, the plot shows that the nonlocal counterterm~\reef{eq:phict} captures most of the $\SO(d+1)$-breaking effects.

The analysis in Sec.~\ref{sec:leador} only considered leading-order contributions to one-point functions. Nonetheless, by examining the two diagrams $\mca{E}_{n,2}$ from~\reef{eq:vevmic}, we can learn something about Ward identity violations at subleading orders. The relevant contributions $\mca{E}_{n,2}(\tau,\La)$ are slightly more complicated. There are two qualitatively different diagrams $\mca{E}^{(1,2)}$ that need to be taken into account, corresponding to different time-orderings:
\bsub
\begin{multline}
  \mca{E}_{n,2}^{(1)}(\tau,\La) = \frac{2n}{n! \Sd }  \sum \Theta(2\vareps(\l) \leq \La)\Theta(\vareps(\l) + \vareps(j_1) + \ldots + \vareps(j_{n-1}) \leq \La) A_n(\l,j_1,\ldots,j_{n-1})  \\
  \times \left[  K_\l(\tau)^2 \int_{-\infty}^\tau\!\frac{d\sigma}{(\cosh \sigma)^d}\int_{-\infty}^\sigma\!\frac{d\sigma'}{(\cosh \sigma')^d}\, K_\l(-\sigma)K_{\l}(-\sigma') \prod_{i=1}^{n-1} K_{j_i}(\sigma)K_{j_i}(-\sigma') + (\tau \to -\tau) \right]
\end{multline}
and
  \begin{multline}
    \label{eq:altn2}
  \mca{E}_{n,2}^{(2)}(\tau,\La) = \frac{2n}{n! \Sd }  \sum \Theta(\vareps(\l) + \vareps(j_1) + \ldots + \vareps(j_{n-1}) \leq \La) A_n(\l,j_1,\ldots,j_{n-1})  \\
  \times \;  K_\l(\tau)K_\l(-\tau) \int_{\tau}^\infty\!\frac{d\sigma}{(\cosh \sigma)^d}\int_{-\infty}^\tau\!\frac{d\sigma'}{(\cosh \sigma')^d}\, K_\l(\sigma)K_{\l}(-\sigma') \prod_{i=1}^{n-1} K_{j_i}(\sigma)K_{j_i}(-\sigma').
\end{multline}
  \esub
  In both expressions, the sum runs over all tuples $\{\l,j_1,\ldots,j_{n-1}\}$. The above expression does not take into account the RG-improvement counterterm~\reef{eq:phi2ct} that is generated at order $\la_3^2$. Adding this counterterm to the action, the function $\mca{E}_{3,2}(\tau,\La)$ shifts as 
\beq
\mca{E}_{3,2}(\tau,\La) \longrightarrow \mca{E}_{3,2}(\tau,\La) + \frac{1}{4\cdot 64\cdot\La R} + O(1/\La^2).
\eeq

Let us discuss the quadratic and cubic interactions separately, starting with the quadratic ($n=2$) one. By numerically evaluating the diagram $\mca{E}_{2,2}$, we find that rotation-invariance violation effects are of the order of $1/\La^3$:
\beq
\mca{E}_{2,2}(\tau,\La) = \frac{1}{8} - \frac{f_2(\tau)}{(\La R)^3} + O(1/\La^4)
\eeq
for some function $f_2(\tau)$. Since truncation error scales as $1/\La^3$, it is negligible  even for moderate values of the cutoff, and we have not attempted to compute the function $f_2(\tau)$.

The convergence of the cubic ($n=3$) term with $\La$ is much slower. A numerical analysis of $\mca{E}_{3,2}$ shows that there is a residual truncation error of order $1/\La$, contrary to the $1/\La^3$ scaling in the quadratic case. In principle, we could further renormalize the operator $\phi^2$ by adding an additional nonlocal counterterm 
\beq
\phi^2_\text{r}(\tau,\bn) \mapsto \phi^2_\text{r}(\tau,\bn) + \frac{\la_3^2 R}{\La} f_3(\tau) \unit
\eeq
to cancel these $1/\La$ rotation-invariance breaking effects. Determining the function $f_3(\tau)$ is however a difficult exercise that will be left for future work.

\section{Discussion}

In this paper we developed a new framework to perform QFT computations on the $d$-dimensional sphere. In the case of the $\phi^2$ flow in $d=3$, we found good agreement between the data computed in our scheme and analytic results. For the $i\phi^3$ flow on $S^3$, we provided strong numerical evidence that the theory is UV-finite after adding a local counterterm that grows logarithmically with the cutoff $\La$. Finally, we analyzed the violation of $\SO(d+1)$ Ward identities in the case of the one-point function $\expec{\phi^2}$ on $S^3$ in perturbation theory, and showed that such violations vanish in the continuum limit $\La \to \infty$.

It is an outstanding problem to compute the $F$-coefficient for nontrivial 3$d$ CFTs, like the Ising or Lee-Yang theories, to high precision. This requires performing a finetuning in the UV (in order to reach the critical point) and computing the partition function $Z(R)$ for a range of radii $R \gg 1$, in order to substract the two curvature counterterms from Eq.~\reef{eq:3dZ}. An estimate of $F$ for the $3d$ Ising model, obtained using the epsilon expansion, was given in~\cite{Giombi:2014xxa}. We are not aware of any published estimate of $F$ for the Lee-Yang CFT. Moreover, it would be interesting to compare integrated correlation functions for the $3d$ Ising model on $S^3$ to recent predictions for Binder cumulants from Ref.~\cite{Berkowitz:2018ssj}.

Although we specialized to scalar quantum field theories on $S^d$, our approach is rather general. It is completely straightforward to extend the same method to theories with fermions. Moreover, the $S^d$ geometry only featured in two places: through the mode functions $K_\l(\tau)$ that appeared in the quantization of $\phi$, and via a factor $1/(\cosh \tau)^d$ appearing in the Dyson operator~\reef{eq:Udef1}. The same method can be adapted to any manifold that is Weyl-equivalent to $\mbb{R} \times S^{d-1}$. Examples of such manifolds are Euclidean anti-de Sitter space (in the Poincar\'{e} disk picture), flat space and de Sitter space. In the case of de Sitter space, the quantization in question was recently used in Ref.~\cite{Kumar:2018iie}, which studied the large-$N$ limit of the $O(N)$ fixed point on dS${}_3$ using gap equations.

The truncation procedure used in this work was chosen more or less {ad hoc}, and it seems worthwhile to examine whether a different truncation might be more efficient. Moreover, we evaluated matrix elements of the Dyson operator $U$ explicitly, by dividing the sphere into $T \gg 1$ timeslices and summing over all intermediate states. This is numerically inexpensive for small cutoffs of order $\La R \sim 10-20$, but the exponential growth of the Fock space with $\La$ means that cutoffs beyond $\La R \sim 30-40$ are not easily accessible. To obtain precise results, it is therefore crucial to compute further RG-improvement counterterms, for instance those of order $1/\La^2$.

We did not use any CFT techniques in our numerical computations, although the UV theory used  was conformally invariant (having $m^2 = 0$). We believe that conformal symmetry could be a powerful tool to improve our approach. For one, it completely fixes the matrix elements $\braket{\psi_i}{\wh{\phi^n}(\tau)}{\psi_j}$ in terms of a small number of OPE coefficients. The next-to-leading term in the expansion of the Dyson operator is an integral over a matrix element of the form
\beq
\label{eq:cft4p}
\braket{\psi_i}{ \wh{\phi^n}(\tau)\wh{\phi^{n'}}(\tau')}{\psi_j}.
\eeq
Such integrals can be estimated accurately by interpreting~\reef{eq:cft4p} as a CFT four-point function, expanding the latter into conformal blocks and summing over all descendants. Taking such CFT ideas into account, it seems possible to increase the cutoff and simultaneously reduce the computational cost (the number $T$ of timeslices).

As a final comment, we mention that in earlier work on so-called spherical or modal field theory a stochastic technique (diffusion Monte Carlo) was used to compute a similar quantity, namely a transition amplitude on $\mbb{R} \times S^1$~\cite{Marrero:1999gf} resp.\@ $\mbb{R} \times  S^2$~\cite{Windoloski:2000yb}. It would be interesting to compare both methods, or perhaps to apply an entirely different numerical approach to our problem.

\subsubsection*{Acknowledgements}

This research was supported in part by Perimeter Institute for Theoretical Physics. Research at Perimeter Institute is supported by the Government of Canada through the Department of Innovation, Science and Economic Development Canada and by the Province of Ontario through the Ministry of Research, Innovation and Science. The author thanks Freddy Cachazo, Jaume Gomis, Niamh Maher, Lorenzo di Pietro, Silviu Pufu, Leonardo Rastelli, Slava Rychkov and Balt van Rees for discussions or comments, and Slava Rychkov for comments on the manuscript.

 {\fns 
\bibliographystyle{utphys}
\bibliography{biblio}

\providecommand{\href}[2]{#2}\begingroup\raggedright\begin{thebibliography}{10}

\bibitem{Zamolodchikov:2001dz}
{\relax Al}.~B. Zamolodchikov, ``{Scaling Lee-Yang model on a sphere. 1.
  Partition function},''
  \href{http://dx.doi.org/10.1088/1126-6708/2002/07/029}{{\em JHEP} {\bfseries
  07} (2002) 029},
\href{http://arxiv.org/abs/hep-th/0109078}{{\ttfamily arXiv:hep-th/0109078
  [hep-th]}}.

\bibitem{Zamolodchikov:1986gt}
A.~Zamolodchikov, ``{Irreversibility of the Flux of the Renormalization Group
  in a 2D Field Theory},''
{\em JETP Lett.} {\bfseries 43} (1986) 730--732.

\bibitem{Cardy:1988cwa}
J.~L. Cardy, ``{Is there a $c$-theorem in four dimensions?},''
\href{http://dx.doi.org/10.1016/0370-2693(88)90054-8}{{\em Phys. Lett.}
  {\bfseries B215} (1988) 749--752}.

\bibitem{Casini:2012ei}
H.~Casini and M.~Huerta, ``{On the RG running of the entanglement entropy of a
  circle},'' \href{http://dx.doi.org/10.1103/PhysRevD.85.125016}{{\em Phys.
  Rev.} {\bfseries D85} (2012) 125016},
\href{http://arxiv.org/abs/1202.5650}{{\ttfamily arXiv:1202.5650 [hep-th]}}.

\bibitem{Liu:2012eea}
H.~Liu and M.~Mezei, ``{A Refinement of entanglement entropy and the number of
  degrees of freedom},'' \href{http://dx.doi.org/10.1007/JHEP04(2013)162}{{\em
  JHEP} {\bfseries 04} (2013) 162},
\href{http://arxiv.org/abs/1202.2070}{{\ttfamily arXiv:1202.2070 [hep-th]}}.

\bibitem{Klebanov:2012va}
I.~R. Klebanov, T.~Nishioka, S.~S. Pufu, and B.~R. Safdi, ``{Is Renormalized
  Entanglement Entropy Stationary at RG Fixed Points?},''
  \href{http://dx.doi.org/10.1007/JHEP10(2012)058}{{\em JHEP} {\bfseries 10}
  (2012) 058},
\href{http://arxiv.org/abs/1207.3360}{{\ttfamily arXiv:1207.3360 [hep-th]}}.

\bibitem{Casini:2015woa}
H.~Casini, M.~Huerta, R.~C. Myers, and A.~Yale, ``{Mutual information and the
  $F$-theorem},'' \href{http://dx.doi.org/10.1007/JHEP10(2015)003}{{\em JHEP}
  {\bfseries 10} (2015) 003},
\href{http://arxiv.org/abs/1506.06195}{{\ttfamily arXiv:1506.06195 [hep-th]}}.

\bibitem{Ben-Ami:2015zsa}
O.~Ben-Ami, D.~Carmi, and M.~Smolkin, ``{Renormalization group flow of
  entanglement entropy on spheres},''
  \href{http://dx.doi.org/10.1007/JHEP08(2015)048}{{\em JHEP} {\bfseries 08}
  (2015) 048},
\href{http://arxiv.org/abs/1504.00913}{{\ttfamily arXiv:1504.00913 [hep-th]}}.

\bibitem{Klebanov:2011gs}
I.~R. Klebanov, S.~S. Pufu, and B.~R. Safdi, ``{$F$-Theorem without
  Supersymmetry},'' \href{http://dx.doi.org/10.1007/JHEP10(2011)038}{{\em JHEP}
  {\bfseries 10} (2011) 038},
\href{http://arxiv.org/abs/1105.4598}{{\ttfamily arXiv:1105.4598 [hep-th]}}.

\bibitem{Anninos:2012ft}
D.~Anninos, F.~Denef, and D.~Harlow, ``{Wave function of Vasiliev’s universe:
  A few slices thereof},''
  \href{http://dx.doi.org/10.1103/PhysRevD.88.084049}{{\em Phys. Rev.}
  {\bfseries D88} no.~8, (2013) 084049},
\href{http://arxiv.org/abs/1207.5517}{{\ttfamily arXiv:1207.5517 [hep-th]}}.

\bibitem{Fei:2014yja}
L.~Fei, S.~Giombi, and I.~R. Klebanov, ``{Critical $O(N)$ models in
  $6-\epsilon$ dimensions},''
  \href{http://dx.doi.org/10.1103/PhysRevD.90.025018}{{\em Phys. Rev.}
  {\bfseries D90} no.~2, (2014) 025018},
\href{http://arxiv.org/abs/1404.1094}{{\ttfamily arXiv:1404.1094 [hep-th]}}.

\bibitem{Giombi:2014xxa}
S.~Giombi and I.~R. Klebanov, ``{Interpolating between $a$ and $F$},''
  \href{http://dx.doi.org/10.1007/JHEP03(2015)117}{{\em JHEP} {\bfseries 03}
  (2015) 117},
\href{http://arxiv.org/abs/1409.1937}{{\ttfamily arXiv:1409.1937 [hep-th]}}.

\bibitem{Fei:2015oha}
L.~Fei, S.~Giombi, I.~R. Klebanov, and G.~Tarnopolsky, ``{Generalized
  $F$-Theorem and the $\epsilon$ Expansion},''
  \href{http://dx.doi.org/10.1007/JHEP12(2015)155}{{\em JHEP} {\bfseries 12}
  (2015) 155},
\href{http://arxiv.org/abs/1507.01960}{{\ttfamily arXiv:1507.01960 [hep-th]}}.

\bibitem{Giombi:2015haa}
S.~Giombi, I.~R. Klebanov, and G.~Tarnopolsky, ``{Conformal QED$_d$,
  $F$-Theorem and the $\epsilon$ Expansion},''
  \href{http://dx.doi.org/10.1088/1751-8113/49/13/135403}{{\em J. Phys.}
  {\bfseries A49} no.~13, (2016) 135403},
\href{http://arxiv.org/abs/1508.06354}{{\ttfamily arXiv:1508.06354 [hep-th]}}.

\bibitem{Tarnopolsky:2016vvd}
G.~Tarnopolsky, ``{Large $N$ expansion of the sphere free energy},''
  \href{http://dx.doi.org/10.1103/PhysRevD.96.025017}{{\em Phys. Rev.}
  {\bfseries D96} no.~2, (2017) 025017},
\href{http://arxiv.org/abs/1609.09113}{{\ttfamily arXiv:1609.09113 [hep-th]}}.

\bibitem{Jafferis:2010un}
D.~L. Jafferis, ``{The Exact Superconformal R-Symmetry Extremizes $Z$},''
  \href{http://dx.doi.org/10.1007/JHEP05(2012)159}{{\em JHEP} {\bfseries 05}
  (2012) 159},
\href{http://arxiv.org/abs/1012.3210}{{\ttfamily arXiv:1012.3210 [hep-th]}}.

\bibitem{Jafferis:2011zi}
D.~L. Jafferis, I.~R. Klebanov, S.~S. Pufu, and B.~R. Safdi, ``{Towards the
  $F$-Theorem: $\mathcal{N}=2$ Field Theories on the Three-Sphere},''
  \href{http://dx.doi.org/10.1007/JHEP06(2011)102}{{\em JHEP} {\bfseries 06}
  (2011) 102},
\href{http://arxiv.org/abs/1103.1181}{{\ttfamily arXiv:1103.1181 [hep-th]}}.

\bibitem{Gulotta:2011si}
D.~R. Gulotta, C.~P. Herzog, and S.~S. Pufu, ``{From Necklace Quivers to the
  $F$-theorem, Operator Counting, and $T(U(N))$},''
  \href{http://dx.doi.org/10.1007/JHEP12(2011)077}{{\em JHEP} {\bfseries 12}
  (2011) 077},
\href{http://arxiv.org/abs/1105.2817}{{\ttfamily arXiv:1105.2817 [hep-th]}}.

\bibitem{Closset:2012vg}
C.~Closset, T.~T. Dumitrescu, G.~Festuccia, Z.~Komargodski, and N.~Seiberg,
  ``{Contact Terms, Unitarity, and $F$-Maximization in Three-Dimensional
  Superconformal Theories},''
  \href{http://dx.doi.org/10.1007/JHEP10(2012)053}{{\em JHEP} {\bfseries 10}
  (2012) 053},
\href{http://arxiv.org/abs/1205.4142}{{\ttfamily arXiv:1205.4142 [hep-th]}}.

\bibitem{ZKnotes}
Z.~Komargodski, ``{Aspects of Renormalization Group Flows}.''
\newblock \url{http://goo.gl/kKdoJb}.

\bibitem{Brower:2016moq}
R.~C. Brower, G.~Fleming, A.~Gasbarro, T.~Raben, C.-I. Tan, and E.~Weinberg,
  ``{Quantum Finite Elements for Lattice Field Theory},'' {\em PoS} {\bfseries
  LATTICE2015} (2016) 296,
\href{http://arxiv.org/abs/1601.01367}{{\ttfamily arXiv:1601.01367 [hep-lat]}}.

\bibitem{Brower:2016vsl}
R.~C. Brower, E.~S. Weinberg, G.~T. Fleming, A.~D. Gasbarro, T.~G. Raben, and
  C.-I. Tan, ``{Lattice Dirac Fermions on a Simplicial Riemannian Manifold},''
  \href{http://dx.doi.org/10.1103/PhysRevD.95.114510}{{\em Phys. Rev.}
  {\bfseries D95} no.~11, (2017) 114510},
\href{http://arxiv.org/abs/1610.08587}{{\ttfamily arXiv:1610.08587 [hep-lat]}}.

\bibitem{Brower:2018szu}
R.~C. Brower, M.~Cheng, E.~S. Weinberg, G.~T. Fleming, A.~D. Gasbarro, T.~G.
  Raben, and C.-I. Tan, ``{Lattice $\phi^4$ field theory on Riemann manifolds:
  Numerical tests for the 2-d Ising CFT on $\mathbb{S}^2$},''
  \href{http://dx.doi.org/10.1103/PhysRevD.98.014502}{{\em Phys. Rev.}
  {\bfseries D98} no.~1, (2018) 014502},
\href{http://arxiv.org/abs/1803.08512}{{\ttfamily arXiv:1803.08512 [hep-lat]}}.

\bibitem{Brower:2012vg}
R.~C. Brower, G.~T. Fleming, and H.~Neuberger, ``{Lattice Radial Quantization:
  3D Ising},'' \href{http://dx.doi.org/10.1016/j.physletb.2013.03.009}{{\em
  Phys. Lett.} {\bfseries B721} (2013) 299--305},
\href{http://arxiv.org/abs/1212.6190}{{\ttfamily arXiv:1212.6190 [hep-lat]}}.

\bibitem{Brower:2012mn}
R.~C. Brower, G.~T. Fleming, and H.~Neuberger, ``{Radial Quantization for
  Conformal Field Theories on the Lattice},'' {\em PoS} {\bfseries LATTICE2012}
  (2012) 061,
\href{http://arxiv.org/abs/1212.1757}{{\ttfamily arXiv:1212.1757 [hep-lat]}}.

\bibitem{Brower:2014daa}
R.~C. Brower, M.~Cheng, and G.~T. Fleming, ``{Improved Lattice Radial
  Quantization},'' {\em PoS} {\bfseries LATTICE2013} (2014) 335,
\href{http://arxiv.org/abs/1407.7597}{{\ttfamily arXiv:1407.7597 [hep-lat]}}.

\bibitem{truncRev}
A.~J.~A. James, R.~M. Konik, P.~Lecheminant, N.~J. Robinson, and A.~M. Tsvelik,
  ``{Non-perturbative methodologies for low-dimensional strongly-correlated
  systems: From non-abelian bosonization to truncated spectrum methods},''
  \href{http://dx.doi.org/10.1088/1361-6633/aa91ea}{{\em Rept. Prog. Phys.}
  {\bfseries 81} no.~4, (2018) 046002},
\href{http://arxiv.org/abs/1703.08421}{{\ttfamily arXiv:1703.08421
  [cond-mat.str-el]}}.

\bibitem{Katz:2013qua}
E.~Katz, G.~Marques~Tavares, and Y.~Xu, ``{Solving 2D QCD with an adjoint
  fermion analytically},''
  \href{http://dx.doi.org/10.1007/JHEP05(2014)143}{{\em JHEP} {\bfseries 05}
  (2014) 143},
\href{http://arxiv.org/abs/1308.4980}{{\ttfamily arXiv:1308.4980 [hep-th]}}.

\bibitem{Hogervorst:2014rta}
M.~Hogervorst, S.~Rychkov, and B.~C. van Rees, ``{Truncated conformal space
  approach in $d$ dimensions: A cheap alternative to lattice field theory?},''
  \href{http://dx.doi.org/10.1103/PhysRevD.91.025005}{{\em Phys. Rev.}
  {\bfseries D91} (2015) 025005},
\href{http://arxiv.org/abs/1409.1581}{{\ttfamily arXiv:1409.1581 [hep-th]}}.

\bibitem{Katz:2014uoa}
E.~Katz, G.~Marques~Tavares, and Y.~Xu, ``{A solution of 2D QCD at Finite $N$
  using a conformal basis},''
\href{http://arxiv.org/abs/1405.6727}{{\ttfamily arXiv:1405.6727 [hep-th]}}.

\bibitem{Rychkov:2014eea}
S.~Rychkov and L.~G. Vitale, ``{Hamiltonian truncation study of the $\phi^4$
  theory in two dimensions},''
  \href{http://dx.doi.org/10.1103/PhysRevD.91.085011}{{\em Phys. Rev.}
  {\bfseries D91} (2015) 085011},
\href{http://arxiv.org/abs/1412.3460}{{\ttfamily arXiv:1412.3460 [hep-th]}}.

\bibitem{Rychkov:2015vap}
S.~Rychkov and L.~G. Vitale, ``{Hamiltonian truncation study of the $\phi^4$
  theory in two dimensions. II. The $\mathbb Z_2$ -broken phase and the Chang
  duality},'' \href{http://dx.doi.org/10.1103/PhysRevD.93.065014}{{\em Phys.
  Rev.} {\bfseries D93} no.~6, (2016) 065014},
\href{http://arxiv.org/abs/1512.00493}{{\ttfamily arXiv:1512.00493 [hep-th]}}.

\bibitem{Elias-Miro:2015bqk}
J.~Elias-Miro, M.~Montull, and M.~Riembau, ``{The renormalized Hamiltonian
  truncation method in the large $E_T$ expansion},''
  \href{http://dx.doi.org/10.1007/JHEP04(2016)144}{{\em JHEP} {\bfseries 04}
  (2016) 144},
\href{http://arxiv.org/abs/1512.05746}{{\ttfamily arXiv:1512.05746 [hep-th]}}.

\bibitem{Katz:2016hxp}
E.~Katz, Z.~U. Khandker, and M.~T. Walters, ``{A Conformal Truncation Framework
  for Infinite-Volume Dynamics},''
  \href{http://dx.doi.org/10.1007/JHEP07(2016)140}{{\em JHEP} {\bfseries 07}
  (2016) 140},
\href{http://arxiv.org/abs/1604.01766}{{\ttfamily arXiv:1604.01766 [hep-th]}}.

\bibitem{Balthazar:2016utu}
B.~Balthazar, V.~A. Rodriguez, and X.~Yin, ``{Hamiltonian Truncation Study of
  Supersymmetric Quantum Mechanics: S-Matrix and Metastable States},''
\href{http://arxiv.org/abs/1610.07275}{{\ttfamily arXiv:1610.07275 [hep-th]}}.

\bibitem{Elias-Miro:2017tup}
J.~Elias-Miro, S.~Rychkov, and L.~G. Vitale, ``{NLO Renormalization in the
  Hamiltonian Truncation},''
  \href{http://dx.doi.org/10.1103/PhysRevD.96.065024}{{\em Phys. Rev.}
  {\bfseries D96} no.~6, (2017) 065024},
\href{http://arxiv.org/abs/1706.09929}{{\ttfamily arXiv:1706.09929 [hep-th]}}.

\bibitem{Elias-Miro:2017xxf}
J.~Elias-Miro, S.~Rychkov, and L.~G. Vitale, ``{High-Precision Calculations in
  Strongly Coupled Quantum Field Theory with Next-to-Leading-Order Renormalized
  Hamiltonian Truncation},''
  \href{http://dx.doi.org/10.1007/JHEP10(2017)213}{{\em JHEP} {\bfseries 10}
  (2017) 213},
\href{http://arxiv.org/abs/1706.06121}{{\ttfamily arXiv:1706.06121 [hep-th]}}.

\bibitem{Whitsitt:2017ocl}
S.~Whitsitt, M.~Schuler, L.-P. Henry, A.~M. L\"{a}uchli, and S.~Sachdev,
  ``{Spectrum of the Wilson-Fisher conformal field theory on the torus},''
  \href{http://dx.doi.org/10.1103/PhysRevB.96.035142}{{\em Phys. Rev.}
  {\bfseries B96} no.~3, (2017) 035142},
\href{http://arxiv.org/abs/1701.03111}{{\ttfamily arXiv:1701.03111
  [cond-mat.str-el]}}.

\bibitem{Anand:2017yij}
N.~Anand, V.~X. Genest, E.~Katz, Z.~U. Khandker, and M.~T. Walters, ``{RG flow
  from $\phi^4$ theory to the 2D Ising model},''
  \href{http://dx.doi.org/10.1007/JHEP08(2017)056}{{\em JHEP} {\bfseries 08}
  (2017) 056},
\href{http://arxiv.org/abs/1704.04500}{{\ttfamily arXiv:1704.04500 [hep-th]}}.

\bibitem{Rutter:2018aog}
D.~Rutter and B.~C. van Rees, ``{Counterterms in Truncated Conformal
  Perturbation Theory},''
\href{http://arxiv.org/abs/1803.05798}{{\ttfamily arXiv:1803.05798 [hep-th]}}.

\bibitem{Fitzpatrick:2018ttk}
A.~L. Fitzpatrick, J.~Kaplan, E.~Katz, L.~G. Vitale, and M.~T. Walters,
  ``{Lightcone effective Hamiltonians and RG flows},''
  \href{http://dx.doi.org/10.1007/JHEP08(2018)120}{{\em JHEP} {\bfseries 08}
  (2018) 120},
\href{http://arxiv.org/abs/1803.10793}{{\ttfamily arXiv:1803.10793 [hep-th]}}.

\bibitem{Bender:2012ea}
C.~M. Bender, V.~Branchina, and E.~Messina, ``{Ordinary versus PT-symmetric
  $\phi^3$ quantum field theory},''
  \href{http://dx.doi.org/10.1103/PhysRevD.85.085001}{{\em Phys. Rev.}
  {\bfseries D85} (2012) 085001},
\href{http://arxiv.org/abs/1201.1244}{{\ttfamily arXiv:1201.1244 [hep-th]}}.

\bibitem{harm}
K.~Atkinson and W.~Han, \href{http://dx.doi.org/10.1007/978-3-642-25983-8}{{\em
  {Spherical Harmonics and Approximations on the Unit Sphere: An
  Introduction}}}.
\newblock Springer, 2012.

\bibitem{Osborn:1999az}
H.~Osborn and G.~M. Shore, ``{Correlation functions of the energy momentum
  tensor on spaces of constant curvature},''
  \href{http://dx.doi.org/10.1016/S0550-3213(99)00775-0}{{\em Nucl. Phys.}
  {\bfseries B571} (2000) 287--357},
\href{http://arxiv.org/abs/hep-th/9909043}{{\ttfamily arXiv:hep-th/9909043
  [hep-th]}}.

\bibitem{Birrell:1982ix}
N.~Birrell and P.~Davies, {\em {Quantum Fields in Curved Space}}.
\newblock Cambridge University Press,
1982.
\newblock

\bibitem{simonb}
B.~Simon, {\em Functional Integration and Quantum Physics}.
\newblock AMS Chelsea Publishing, 2005.

\bibitem{zamolBook}
{\relax Al}.~B. Zamolodchikov,
  \href{http://dx.doi.org/10.1007/978-94-010-0514-2_10}{``Perturbed conformal
  field theory on a sphere,''} in {\em Statistical Field Theories}, A.~Cappelli
  and G.~Mussardo, eds., pp.~105--116.
\newblock Springer, 2002.

\bibitem{Grinza:2003ji}
P.~Grinza and N.~Magnoli, ``{On the magnetic perturbation of the Ising model on
  the sphere},'' \href{http://dx.doi.org/10.1088/0305-4470/36/39/102}{{\em J.
  Phys.} {\bfseries A36} (2003) L509--L516},
\href{http://arxiv.org/abs/hep-th/0306100}{{\ttfamily arXiv:hep-th/0306100
  [hep-th]}}.

\bibitem{Rychkov:2016iqz}
S.~Rychkov, \href{http://dx.doi.org/10.1007/978-3-319-43626-5}{{\em {EPFL
  Lectures on Conformal Field Theory in $D \geq 3$ Dimensions}}}.
\newblock SpringerBriefs in Physics. 2016.
\newblock
\href{http://arxiv.org/abs/1601.05000}{{\ttfamily arXiv:1601.05000 [hep-th]}}.
\newblock

\bibitem{Simmons-Duffin:2016gjk}
D.~Simmons-Duffin, \href{http://dx.doi.org/10.1142/9789813149441_0001}{``{The
  Conformal Bootstrap},''} in {\em {Proceedings, Theoretical Advanced Study
  Institute in Elementary Particle Physics: New Frontiers in Fields and Strings
  (TASI 2015): Boulder, CO, USA, June 1-26, 2015}}, pp.~1--74.
\newblock 2017.
\newblock
\href{http://arxiv.org/abs/1602.07982}{{\ttfamily arXiv:1602.07982 [hep-th]}}.
\newblock

\bibitem{Symanzik:1983dc}
K.~Symanzik, ``{Continuum Limit and Improved Action in Lattice Theories. 1.
  Principles and $\phi^4$ Theory},''
\href{http://dx.doi.org/10.1016/0550-3213(83)90468-6}{{\em Nucl.Phys.}
  {\bfseries B226} (1983) 187}.

\bibitem{Giokas:2011ix}
P.~Giokas and G.~Watts, ``{The renormalisation group for the truncated
  conformal space approach on the cylinder},''
\href{http://arxiv.org/abs/1106.2448}{{\ttfamily arXiv:1106.2448 [hep-th]}}.

\bibitem{Ghosh:2018qtg}
J.~K. Ghosh, E.~Kiritsis, F.~Nitti, and L.~T. Witkowski, ``{Holographic RG
  flows on curved manifolds and the $F$-theorem},''
\href{http://arxiv.org/abs/1810.12318}{{\ttfamily arXiv:1810.12318 [hep-th]}}.

\bibitem{Berkowitz:2018ssj}
D.~Berkowitz, ``{Conformal invariance and the Ising model on a 3 sphere in
  connection with the Quantum Elemental Method},''
\href{http://arxiv.org/abs/1808.05862}{{\ttfamily arXiv:1808.05862 [hep-th]}}.

\bibitem{Kumar:2018iie}
S.~P. Kumar and V.~Vaganov, ``{Nonequilibrium dynamics of the $O(N)$ model on
  dS$_{3}$ and AdS crunches},''
  \href{http://dx.doi.org/10.1007/JHEP03(2018)092}{{\em JHEP} {\bfseries 03}
  (2018) 092},
\href{http://arxiv.org/abs/1802.08202}{{\ttfamily arXiv:1802.08202 [hep-th]}}.

\bibitem{Marrero:1999gf}
P.~J. Marrero, E.~A. Roura, and D.~Lee, ``{A nonperturbative analysis of
  symmetry breaking in two-dimensional $\phi^4$ theory using periodic field
  methods},'' \href{http://dx.doi.org/10.1016/S0370-2693(99)01341-6}{{\em Phys.
  Lett.} {\bfseries B471} (1999) 45},
\href{http://arxiv.org/abs/hep-th/9906189}{{\ttfamily arXiv:hep-th/9906189
  [hep-th]}}.

\bibitem{Windoloski:2000yb}
M.~Windoloski, ``{A Nonperturbative study of three-dimensional $\phi^4$
  theory},''
\href{http://arxiv.org/abs/hep-th/0002243}{{\ttfamily arXiv:hep-th/0002243
  [hep-th]}}.

\bibitem{FS}
J.~Fuchs and C.~Schweigert, {\em {Symmetries, Lie Algebras and
  Representations}}.
\newblock Cambridge University Press, 2003.

\end{thebibliography}\endgroup
 }

 \appendix

 \section{Spherical harmonics and Gegenbauer polynomials}
\label{sec:harmonics}

We will employ the usual normalization for the spherical harmonics:
\beq
\label{eq:Ynorm}
\int_{S^{d-1}}\!d\bn \, Y^*_{\l j}(\bn) Y^\phs_{\l'j'}(\bn) = \dd_{\l \l'} \dd_{j j'}\,.
\eeq
As is clear from~\reef{eq:Ynorm}, the harmonics corresponding to a given spin $\l$ are only fixed up to a unitary change of basis:
\beq
Y_{\l j}(\bn) \, \mapsto \, \sum_{j'} M_j{}^{j'} Y_{\l j'}(\bn),
\quad
M^\dagger = M^{-1}\,.
\eeq
Such a change of basis does not influence any physical results.

An important role will be played by the Gegenbauer polynomials $C_\l^d(z)$, which we normalize such that $C_\l^d(1) = 1$. Concretely, the $C_\l^d(z)$ can be defined using a generating function
\beq
\frac{1}{(1-2zt+t^2)^{\half d-1}} = \sum_{\l=0}^\infty t^\l \frac{(d-2)_\l}{\l!}\, C_\l^d(z)
\eeq
and they are orthonormal in the following sense:
\beq
\label{eq:Gegn}
 \frac{n_{\l}^d \cdot \mrm{S}_{d-1}}{\Sd} \int_{-1}^1\!dz\, (1-z^2)^{\half(d-3)} C_{\l}^d(z) C_{\l'}^d(z) = \dd_{\l\l'}.
 \eeq
The spherical harmonics are related to Gegenbauer polynomials via the so-called addition theorem:
\beq
\label{eq:addition}
\sum_{j=1}^{n_\l^d} Y^*_{\l j}(\bm) Y^\phs_{\l j}(\bn) = \frac{n_{\l}^d}{\Sd}\,  C_\l^d(\bm \cdot \bn)\,.
\eeq
Although the $Y_{\l j}(\bn)$ appearing on the LHS depend on a choice of basis, the RHS does not. Note that in $d=3$, we simply have $C_\l^3(z) = \mtt{LegendreP}_\l(z)$.

Next, let us consider the computation of various spherical integrals. A fundamental identity, which  slightly generalizes Eq.~\reef{eq:Gegn}, reads
 \beq
 \label{eq:Zam}
 \frac{n_\l^d}{\Sd} \int_{S^{d-1}}\!d{\bf m}\, C^d_{\l}({\bf m}\cdot \bn_1)C^d_{\l'}({\bf m}\cdot \bn_2) = \dd_{\l \l'} \, C^d_{\l}(\bn_1 \cdot \bn_2).
 \eeq
 This can be proven either via~\reef{eq:addition} or by means of the identity
 \beq
 \label{eq:gens}
 \int_{S^{d-1}}\!d\bn\, f(\mathbf{m}\cdot\bn) = \mrm{S}_{d-1} \int_{-1}^1\!dz\, (1-z^2)^{\half(d-3)}f(z),
 \quad
|\mathbf{m}| = 1.
 \eeq
 Next, let us consider the $A_n(\l_1,\ldots,\l_n)$ integrals, defined in Eq.~\reef{eq:Andef}. For $n=2$ the result follows immediately from~\reef{eq:Zam}, namely
 \beq
 A_2(\l_1,\l_2) = \dd_{\l_1\l_2} n_\l^d.
 \eeq
 For $n=3$, we first rewrite the integral as
\beq
\label{eq:A3c}
{A}_3(\l_1,\l_2,\l_3) = n_{\l_1}^d n_{\l_2}^d n_{\l_3}^d \, \frac{\mrm{S}_{d-1}}{\mrm{S}_d^2} \! \int_{-1}^1\!dx \, (1-x^2)^{\half(d-3)}  C_{\l_1}^d(x) C_{\l_2}^d(x) C^d_{\l_3}(x)
\eeq
using~\reef{eq:gens}. The above integral vanishes either if $\l_1 + \l_2 + \l_3$ is odd, or if the triplet $\{\l_1,\l_2,\l_3\}$ does not obey the triangle inequality. If both conditions are fulfilled, the result can be stated as
\beq
\label{eq:A3crj}
 {A}_3(\l_1,\l_2,\l_3) =  \frac{(d-2)_\lambda}{\mrm{S}_d (\th d)_\lambda (\th d-1)^3} \prod_{i=1}^3 (\l_i+\th d-1) \frac{(\th d-1)_{\lambda-\l_i}}{(\lambda-\l_i)!},
 \quad
\lambda = \half(\l_1 + \l_2 + \l_3).
\eeq

\section{Massless integrals}

In the massless limit, the $K_\l(\tau)$ functions simplify in the following way:
\beq
K_\l(\tau) = \frac{1}{\sqrt{2\l+d-2}}\;  e^{-(\l+\half d-1)\tau} (\cosh \tau)^{\half d-1} + O(m^2R^2).
\eeq
This allows for the exact computation of various quantities in the limit $m^2 \to 0$, which we make use of in Sec.~\ref{sec:cutoff}.  As a starting point, let's consider the integral
\bsub
\label{eq:G1}
\begin{align}
\Re(\a+\b) > 0: \quad G_0(\a,\b|\tau) &\ldef \int_{-\infty}^\tau\!\frac{d\tau'}{(\cosh \tau')^\a} \, e^{-\b(\tau-\tau')}\\
\label{eq:G1sol}
&= \frac{\Gamma\big(\th(\a+\b)\big)}{2(\cosh \tau)^\a}\;  {}_2\widetilde{F}_1\!\left[{{1,\a}~\atop~{1+\th(\a+\b)}} \; \Bigg| \; \frac{e^\tau}{2\cosh \tau} \right]
\end{align}
\esub
where ${}_p\wt{F}_q$ denotes the regularized hypergeometric function. 
To prove the above identity, one can e.g.\@ notice that~\reef{eq:G1sol} is the unique solution to the first-order differential equation
\[
\left[ \beta + \frac{d}{d\tau} \right]\!G_0(\a,\b|\tau) = \frac{1}{(\cosh \tau)^\a}
\]
which satisfies $\lim_{\tau \to -\infty} G_0(\a,\b|\tau) = 0$. At large $\beta$, keeping $\a$ and $\tau$ fixed, we have
\beq
\label{eq:G1as}
G_0(\a,\b|\tau) \;\limu{\b \to \infty} \; \frac{1}{(\cosh \tau)^\a \b} + O(1/\b^2).
\eeq
A similar integral is
\beq
\label{eq:G1alt}
\wh{G}_0(\a,\b|\tau) \ldef \int_{-\infty}^\infty \frac{d\tau'}{(\cosh \tau')^\a} \, e^{-\b|\tau-\tau'|} = G_0(\a,\b|\tau) + G_0(\a,\b|\!-\!\tau).
\eeq
For second-order corrections to the partition function, we need
\bsub
\label{eq:G2}
\beq
G_1(\a,\b) \ldef  \int_{-\infty}^\infty\!\frac{d\tau}{(\cosh \tau)^\a} \int_{-\infty}^\tau\!\frac{d\tau'}{(\cosh \tau')^\a} \, e^{-\b(\tau-\tau')} 
\eeq
which converges if $\Re(\a+\b) >0$ as well as $\Re(\a) > 0$. Using Eq.~\reef{eq:G1sol}, it can be shown that the above integral evaluates to
\beq
\label{eq:G2sol}
G_1(\a,\b) = 4^{\a-1} \Gamma(\a)^2 \Gamma\big(\th(\a+\b)\big) \, {}_3\widetilde{F}_2\!\left[{{1,\a,\a}~\atop~{2\a,1+\th(\a+\b)}} \; \Bigg| \; 1\, \right].
\eeq
\esub
In the limit where $\b \gg 1$ but $\a$ is kept fixed, we have
\beq
\label{eq:G2as}
G_1(\a,\b) \limu{\b \to \infty} \beta^{-1} B(\a,\th) + O(1/\b^2).
\eeq

\section{Additional comments about the algorithm}\label{sec:remarks}

In Sec.~\ref{sec:implementation}, a brief description of the algorithm used in our work was given. In what follows, we will discuss three additional points that may be helpful for readers who want to implement a version of the algorithm themselves.

Our first comment involves generating a list of scalar states $\ket{\psi_i}$.  Such states can be written as linear combinations of parity-even states with $L_z = 0$, which can be denoted as $\ket{\chi_\a}$. Schematically
\beq
\label{eq:rel}
\ket{\psi_i} = \sum_\a C\du{i}{\a} \ket{\chi_\a}
\eeq
for some matrix $C$ with real-valued matrix elements. In principle, the matrix $C$ can be computed by requiring that $L_\pm \ket{\psi_i} = 0$ --- it's the kernel of either of the operators $L_+$ or $L_{-}$. Alternatively, the matrix $C\du{i}{\a}$ can be constructed using $\SO(3)$ Lie algebra techniques. Working in a basis of states $\ket{\chi_\a}$ defined in Eq.~\reef{eq:basisState}, every line of the matrix $C\du{i}{\a}$ has a group-theoretical interpretation as a tensor $T$ in a tensor product of $\SO(3)$ representations. We can therefore generate scalar states by writing down manifestly invariant tensors, using Clebsch-Gordan coefficients and 3$j$ symbols~\cite{FS}. For instance, given three spins $\l_i$ that obey the triangle equality and sum to an even integer, there is a unique scalar state:
\beq
\sum_{m_i} \threej{\l_1}{\l_2}{\l_3}{m_1}{m_2}{m_3} a_{\l_1 m_1}^\dagger  a_{\l_2 m_2}^\dagger a_{\l_3 m_3}^\dagger \ket{\emptyset}.
\eeq
The generalization to $n$-particle states is straightforward. In our experience, such a group-theoretical approach is faster than computing the kernel of $L_\pm$.

Second, we proceed by computing the matrix elements of $\wh{\phi_n}(\tau)$ in a basis of the $\ket{\chi_\a}$ states and restricting to scalar states later on. For simplicity, let's work in a basis where $\brakket{\psi_i}{\psi_j} = \dd_{ij}$ and $\brakket{\chi_\a}{\chi_\b} = \dd_{\a\b}$, which implies that
\beq
C \cdot {}^tC = \unit. 
\eeq
Although it is easy to orthonormalize the states $\ket{\chi_\a}$, making the scalars $\ket{\psi_i}$ orthonormal requires the use of the Gram-Schmidt procedure, which is somewhat expensive. Let 
\beq
[W_n(\tau)]\ud{\a}{\b} = [W_n(\tau)]_{\a\b} \ldef \braket{\chi_\a}{\wh{\phi^n}(\tau)}{\chi_\b}.
\eeq
Using~\reef{eq:rel} it follows that
\beq
[V_n(\tau)]\ud{i}{j} = [V_n(\tau)]_{ij} = [C \cdot W_n(\tau) \cdot {}^t C]_{ij}.
\eeq
The matrices $W_n(\tau)$ are rather large --- for $\La R = 20$, they have $166802 \times 166802$ entries, see Table~\ref{table:count}. Using an additional trick, the dimension of $W_n$ can be reduced, as is explained in the final paragraph of this section. Next, notice that hermiticity puts constraints on the matrices $V_n$ and $W_n$, namely
\beq
[V_n(\tau)]_{ij} = [V_n(-\tau)]_{ji}
\qaq
[W_n(\tau)]_{\a\b} = [W_n(-\tau)]_{\b\a}.
\eeq
A second consistency condition follows from rotation invariance. Let $\mca{P}$ be the operator that projects onto scalar states:
\beq
\mca{P} = \sum_{\a\b} \, [{}^t C \cdot C]^{\a\b} \, \ket{\chi_\a}\bra{\chi_\b},
\quad
\mca{P}^2 = \mca{P}.
\eeq
The rotation invariance of $\wh{\phi^n}(\tau)$ implies that
\beq
\label{eq:we}
 \forall i:\quad \mca{P}  \wh{\phi^n}(\tau) \ket{\psi_i} = \wh{\phi^n}(\tau) \ket{\psi_i},
\eeq
which translates to the matrix constraint
\beq
\label{eq:consis}
{}^tC \cdot V_n(\tau) = W_n(\tau) \cdot {}^tC.
\eeq
If Eq.~\reef{eq:consis} is not satisfied, at least one matrix element of $W_n(\tau)$ or $V_n(\tau)$ must be incorrect.

Finally, we point out a trick that can be used to drastically simplify the computation of matrix elements. The key idea is that a basis state $\ket{\chi_\a}$ can always be written in the following form:
\beq
\ket{\chi_\a} = \frac{1}{\sqrt{k_\a!}} \, (a_{0,0}^\dag)^{k_\a} \ket{\xi_\a}
\eeq
for some integer $k_\a \geq 0$, where $\ket{\xi_\a}$ is a state without $a_{0,0}^\dag$ creation operators. We claim that the matrix $W_n(\tau)$ can be expressed in terms of matrix elements of the form $\braket{\xi_\a}{\wh{\phi^{n'}}(\tau)}{\xi_\b}$ for $1 \leq n' \leq n$. This is a simple consequence of the  operator identity
\beq
\label{eq:magic}
[\wh{\phi^n}(\tau),a_{0,0}^\dag] = \frac{n}{\sqrt{\Sd}} \, K_0(\tau) \, \wh{\phi^{n-1}}(\tau),
\eeq
using the convention  $\wh{\phi^0}(\tau) = \Sd \unit$. The proof of Eq.~\reef{eq:magic} is left to the reader.

\end{document}